\documentclass{article}

\pdfoutput=1
\usepackage{amsmath}
\usepackage{jheppub}
\usepackage{graphicx}
\usepackage{slashed}
 \usepackage{url}
\usepackage{amsmath}
\usepackage{amssymb}
\usepackage{amsthm}
\usepackage{verbatim,graphicx}
\usepackage{mathtools}
\usepackage{color}
\usepackage[all]{xy}
\usepackage{simplewick}
\usepackage{hyperref}
\usepackage{setspace}
\usepackage{array}
\usepackage{tikz}
\usepackage{tkz-euclide}
\usepackage{enumitem}
\usepackage{bbm}
\usepackage{bbold}

\newcounter{yjcc}

\newcommand{\Tr}{{\rm Tr}}

\newcommand{\nn}{\nonumber}

\newcommand{\be}{\begin{eqnarray}}
\newcommand{\ee}{\end{eqnarray}}
\newcommand{\vev}[1]{\left\langle #1\right\rangle}

\newcommand{\bmat}{\left ( \begin{array}{cc} }
	\newcommand{\emat}{\end{array} \right ) }
\newcommand{\floor}[1]{\lfloor  #1 \rfloor}

\usetikzlibrary{shapes.misc}
\tikzset{cross/.style={cross out, draw=black, fill=none, minimum size=2*(#1-\pgflinewidth), inner sep=0pt, outer sep=0pt}, cross/.default={2pt}}

\def\Tr{\textrm{Tr}}

\allowdisplaybreaks
\title{Bosonic near-CFT$_1$ models from Fock-space fluxes}
\author{Yiyang Jia }
\affiliation{Department of Particle Physics and Astrophysics, Weizmann Institute of Science, Rehovot, Israel}
\emailAdd{yiyang.jia@weizmann.ac.il}
\date{}
\abstract{ We construct a family of near-CFT$_1$ models with a conserved U(1) charge, whose basic degrees of freedom are canonical bosons. The Sachdev-Ye-Kitaev (SYK) model---the first  microscopic model that realizes the near-CFT$_1$ dynamics---is based on random $p$-local interactions among fermions.  However, a bosonic near-CFT$_1$ model has remained elusive in the $p$-local approach because such constructions generally suffer from unwanted orderings at low temperatures.  Our construction is based on a recent insight that near-CFT$_1$ dynamics can quite generally arise if we place a large amount of random fluxes in a many-body Fock space and $p$-locality is not essential. All such models are essentially solved by chord diagrams regardless of the nature of the underlying degrees of freedom.  We further  argue that such bosonic models do not suffer from energetic instablities  or unwanted low-temperature orderings.    
 For comparison we also consider a second class of charge-conserving models which are based on qubits. The thermodynamic scalings of these models are very similar to those of the double-scaled complex SYK model but are free of certain singularities the latter suffers from.  We also show the level statistics of both models are described by random matrix theory universality down to very low energies.}
\begin{document}
\maketitle
\section{Introduction}
The microscopic dynamics of quantum black holes present an outstanding challenge in our modern understanding of quantum gravity.  Considerable progress has been made for near-extremal black holes due to the advent of the Sachdev-Ye-Kiaev (SYK) model \cite{kitaev2015,sachdev1993,sachdev2010}.  The SYK model is a system of $N$ Majorana fermions interacting through a random $p$-body coupling where each fermion can couple to any of the rest, and $p$ is an integer parametrically smaller than $N$. An interaction of this type is called $p$-local.  The SYK model's relevance to holographic duality is 
realized by the fact that its infrared (IR) dynamics is described by a one-dimensional nearly conformal field theory (NCFT$_1$). This NCFT$_1$ is described by a universal Schwarzian effective action \cite{kitaev2015,maldacena2016} which is dual to  Jackiw-Teitolboim  (JT) gravity. The JT action is the effective theory of the near-horizon dynamics of near-extremal black holes, whose geometry is a two-dimensional nearly anti-de Sitter space (NAdS$_2$) \cite{jackiw1985,teitelboim1983,almheiri2015,maldacena2016a}. 
The SYK model is also historically important as a model for quantum spin liquid \cite{sachdev1993,Georges2000,Georges2001} and as a model for heavy nuclei \cite{french1970,french1971,bohigas1971,bohigas1971a,mon1975}, and is also a useful model for studying many-body quantum chaos.   

The NCFT$_1$/NAdS$_2$ duality in the above sense should be viewed as a universal but nontrivial sector of full-fledged holographic dualities (such as $\mathcal{N}=4$ super Yang-Mills/Type IIB string). In addition, the universal NCFT$_1$ phenomenologies are often robust against certain changes in the microscopic definitions of the SYK model.  For example, we can use any even number $p$ for the $p$-fermion interaction and take the large $N$ limit; we can  take a large $p$ limit after the large $N$ limit;  we can take a double-scaled (DS) limit where $p \sim \sqrt{N}$ as the large $N$ limit is taken.  All the options essentially produce the same universal IR dynamics captured by the Schwarzian action.  Moreover, it is worth noting that if we take the DS limit, we can even replace the fermions by Pauli spins and still get the same IR dynamics \cite{Berkooz:2018qkz,erdos2014}: both the DS-SYK and the DS spin models have the same combinatorial solution described by chord diagrams.  More recently, it is realized that an old model of Parisi \cite{Parisi:1994jg} which is not at all of the form of a random $p$-local interaction, also has the same chord diagram solution and thus the same IR dynamics \cite{Jia_2020,berkooz2023parisis,berkooz2023parisislong}.  Therefore, it seems plausible that there are certain universal aspects of the microscopics that ensure the universal NCFT$_1$ phenomenologies.   A tentative characterization of such universal microscopics is recently given in \cite{berkooz2023parisis,berkooz2023parisislong}, which can be colloquially summarized as the following: one has an  NCFT$_1$ in the IR if in the microscopic Fock space there is a large amount of time-independent, random and uniform fluxes. We will see the precise meaning of it in the following sections. Here let us just mention the effects of such fluxes at two important time scales:  at early times, the constraints we put on the fluxes ensure that the dynamics are captured by chord diagram solutions and therefore exhibit Schwarzian/NCFT$_1$ physics;  at late times the fluxes will sufficiently delocalize the wavefunctions  and therefore the system eigenstate-thermalizes.

 In this paper we will build a few charge-conserving bosonic NCFT$_1$ models using such Fock-space fluxes. In particular, we wish to highlight the case where the basic degrees of freedoms are canonical bosons. A  bosonic NCFT$_1$ has been hard to achieve using $p$-local constructions because they generically have low-temperature orderings such as spin-glass ordering \cite{Georges2000,Georges2001, baldwin2020,Anous_2021, swingle2023bosonic}, so that even though  SYK-like solutions exist but they are unphysical. 
Moreover, for the cases with supersymmetries, the best studied NCFT$_1$ model is where bosons are realized as the composites of the underlying fermions \cite{fu2018,Heydeman_2023}, but UV-complete holographic CFTs---from which the supersymmetric NCFT$_1$ presumably flows from---often has elementary canonical bosons.  Therefore it is interesting to construct supersymmetric (SUSY) NCFT$_1$ models that contain such bosons. There are many interesting developments along this line \cite{Anninos_2016, biggs2023supersymmetric,Benini_2023}, and more recently \cite{benini2024mathcaln2} found an intriguing class of $p$-local SUSY models whose annealed solutions  exhibit  NCFT$_1$ properties and show some good hints of the absence of spin-glass orderings. However,  it is generally a difficult task to make sure of the absence of orderings at low temperatures, especially in a purely bosonic model.
We will argue that our construction of bosonic models does not suffer from the ordering problem for a fairly general reason, and the argument will be made particularly sharp (and easy) for certain representatives of this class of models. Therefore they are valid examples of bosonic NCFT$_1$ models.  The gist is that fluxes suppress the return amplitudes of wavefunctions,  and randomness ensures that there are no constructive interferences when summing up the return amplitudes. And when there are sufficient amount of such fluxes,  the eigenstates become very delocalized, and hence cannot support any ordering. This mechanism also leads to eigenstate thermalization, characterized by the level statistics of the system following random matrix universality, which we will also demonstrate. 

The paper is organized as the following: in section \ref{sec:genTheo} we present the general theory of Fock-space fluxes and how they give rise to an NCFT$_1$ in the IR. The central technique involved is chord diagram combinatorics. We will  also show how various known models, including the DS-SYK model, fall into this category and give a preview of the models that will be studied in this paper.  In section \ref{sec:bosonicModel} we construct the canoncial bosonic model with a conserved particle number, and solve for its thermodynamics and correlation functions in detail using chord diagram techniques. We further demonstrate such bosonic models do not  suffer from low-temperature orderings or energetic instablities. In section \ref{sec:qubitModel}, we study a qubit-based model.  The techniques involved in solving it are very similar to those of the canonical boson models so only the key results are presented.

\section{The general theory}\label{sec:genTheo}
In this section we review the general theory of Fock-space fluxes and its relation to NCFT$_1$, and the logic underlying the constructions of the new models we will study.  However, readers who are keen to see the new models first can jump to section \ref{sec:bosonicModel}.
\subsection{Fluxes in Fock spaces}
Consider a Fock space that is generated by $N$ creation operators $c_i^\dagger (i=1,2,\ldots N)$ , which would describe many-body systems of size $N$. The Fock space is equipped with  an operator
\begin{equation}\label{eqn:chargeOpDef}
    R  = \sum_{i=1}^N r_i,
\end{equation}
where 
\begin{equation} \label{eqn:localChargeOp}
    [r_i, c_j^\dagger] = \delta_{ij}c_j^\dagger, \quad     [r_i, c_j] = -\delta_{ij}c_j,
\end{equation}
so we can call $r_i$ the local occupation number/charge operators on site $i$. We then have $R$ as the total number/charge operator: 
\begin{equation}
       [R, c_j^\dagger] = c_j^\dagger, \quad     [R, c_j] = -c_j.
\end{equation}
To  interpret  it as a space of particles and antiparticles,  one needs to take $N/2$ of the $c_i^\dagger$'s and interpret them as anti-particle annihilation operators, and then $R$ is the charge operator upon normal ordering. Otherwise, $R$ is a number operator.  Later we shall repeatedly make use of the finite form of equation \eqref{eqn:localChargeOp}:
\begin{equation}\label{eqn:chargeOpFiniteTransform}
e^{i u r_i} c_j^\dagger e^{-i u r_i}  = e^{iu \delta_{ij}} c_j^\dagger, \quad e^{i u r_i} c_j  e^{-i u r_i}  = e^{-iu \delta_{ij}} c_j.
\end{equation}
At this point we do not need to specify the nature of the creation/annihilation operators: they can be bosonic, fermionic, qubitized/hardcore-bosonic or perhaps something else. A many-body Hamiltonian in this Fock space is a function $H(\{c_i^\dagger\},\{c_i\})$ of the creation and annihilation operators. 
We can give  a graphical interpretation of the Hamiltonian in the following manner: first represent each Fock basis vector as a point $x$ (whose coordinates are the local occupation numbers $r_i$), the number of points is the Hilbert space dimension which is exponential in $N$. Then connect two points $x_1$ and $x_2$ by a line if and only if the corresponding states give a nonvanishing matrix element for $H$, namely when $\langle x_2| H | x_1 \rangle \neq 0$. 
The graph thus obtained is called a Fock-space graph \cite{altshuler1997} and the graph node degrees should typically diverge in the thermodynamic limit. In this manner, the evolution of a many-body wavefunction is represented by a single particle hopping on an infinite-dimensional graph.  The lines connecting basis points correspond to hopping terms in the Hamiltonian. For example, a hopping term can be a monomial of the form
\begin{equation}
    C_I = c_{i_1}^\dagger \cdots c_{i_{m}}^\dagger c_{i_{m+1}} \cdots c_{i_{m+n}}.
\end{equation}
We can think of the collection of the graded subscripts 
\begin{equation}
    I := \{i_1^+, \ldots, i_m^+, i_{m+1}^-,\ldots, i_{m+n}^-\}
\end{equation}
 as specifying a hopping direction on the Fock-space graph. By Fock-space fluxes we mean fluxes defined on the Fock space graphs, in other words nontrivial holonomies obtained by hoppings along closed loops on the graph.  This definition is somewhat empty so far, in that given any many-body Hamitonian we can always hop along closed loops on its Fock-space graph, and as long as the Hamiltonian is not made of all-commuting terms, there will be some nontrivial holonomies.  However the physics will become highly constrained after we place the following  requirements on fluxes \cite{berkooz2023parisis,berkooz2023parisislong}: 
\[\textit{ The Fock-space fluxes must be uniform, random, and there must be a large amount of them. 
}\]
 We will spell out the precise meaning of these requirements in the following paragraphs.

One convenient way of building highly fluxed Fock-space models is to dress the $C_I$ operators (or some linear combinations of them) with enough fluxes so that they become highly fluxed operators $M_I$, and build the Hamiltonian as\footnote{More generally a Hamiltonian can contain on-site terms $C_I^\dagger C_I$, we work with cases where their effects are negligible.}
\begin{equation}\label{eqn:generalHami}
    H = \sum_I (M_I + M_I^\dagger).
\end{equation}
To build charge-conserving models, one can take those $M_I$ which are charge-neutral under $R$ (defined in equation \eqref{eqn:chargeOpDef}).
The fluxed operators $M_I$ satisfy the algebra
\begin{equation}\label{eqn:magneticAlgebra}
 M_I M_K = e^{i F_{IK}} M_K M_I, \quad M_I M_K^\dagger = e^{-i F_{IK}} M_K^\dagger  M_I.
\end{equation}
The  $F_{IK}$ are the fluxes,  which form an antisymmetric (in $IK$) array of time-independent random numbers (i.e., quench-disordered). The fact that $F_{IK}$ are numbers, as opposed to being a function of the local occupation number operators (which is a more general possibility), means that we should think of them as  being uniform.  Namely, this implies that holonomies are invariant under hoppings:\footnote{In fact we only need this to be true at leading order with probability one ,   for example we are allowed to break the uniformity at orders suppressed by $1/N$, or violate uniformity for a finite number of subscripts.}
\begin{equation}\label{eqn:approxUnifCond2}
    [M_L,  M_K  M_I M_K^\dagger  M_I^\dagger] = 0.
\end{equation}
On the Fock-space graph, this means the phase of a loop only depends on the sequence of hopping directions involved, but does not depend on its  location on the graph. 

 The algebra of \eqref{eqn:magneticAlgebra} is very similar to that of the magnetic translation operators in a lattice Landau problem. As we shall see, here the essential difference lies in the fact that the number of distinct hopping directions goes to infinity as the thermodynamic limit is taken, and in that the fluxes are sufficiently random. Specifically, this allows us to use chord diagram techniques to solve for models in this class. We further assume the fluxes are identically and independently distributed (i.i.d.) for distinct pairs of subscripts,  and moreover 
\begin{equation}\label{eqn:randomFluxCond}
      \vev{\sin F_{IK}} =0, \quad \vev{\cos F_{IK}} =q 
\end{equation}
where $\vev{\cdots}$ denotes ensemble averaging and $q$ is a tunable parameter in the thermodynamic limit. Ultimately we are interested in the $\langle F^2\rangle \to 0$ limit (after the thermodynamic limit is taken), so we may alternatively state the above condition as that the fluxes have a vanishing first moment and  tunable second moment. Note what this really says is that the amount of  fluxes is huge and they are maximally random: for any given pair of indices $I$ and $K$ (and each index can take infinitely many values), the flux
is generically nonzero, and moreover they are independently random among themselves. It is impossible to support this amount of random fluxes using local models such as spin chains.

A few paragraphs ago we gave a colloquial summary of the requirements we impose on the Fock-space fluxes: uniform, random  and a large amount.  By now we have laid out the precise definitions for each of them, which were first articulated in \cite{berkooz2023parisis,berkooz2023parisislong}.  To recapitulate, the meaning of uniform flux is given by equation \eqref{eqn:magneticAlgebra} (which implies equation \eqref{eqn:approxUnifCond2}); random and a large amount are defined by the i.i.d. requirement and equation \eqref{eqn:randomFluxCond}.  These conditions are sufficient to ensure  the universal NAdS$_2$/NCFT$_1$ dynamics as we shall briefly explain.
		The moments of such highly-fluxed models all have the schematic form
\begin{equation}\label{eqn:chordSchematic}
    \vev{\Tr H^{k}} \propto \text{dim} (\mathcal{H}) \sum_{\substack{\text{chord diagrams}\\ \text{with $k/2$ chords}}} q^{\text{number of chord intersections}}, \quad \text{for $k\ll N$},
\end{equation}
where $k$ is even  (we focus on models where odd moments vanish).  We will explain more details of this expression in the coming sections when we use it to solve for the new models of this paper.

One result we shall repeatedly use is that the partition function resulted from summing over the expression on the right-hand side of  \eqref{eqn:chordSchematic} has the following limiting form \cite{Berkooz:2018qkz}
\begin{equation}\label{eqn:dssykParitionFunc}
Z_\text{DSSYK}(\beta)= \text{dim} (\mathcal{H}_\text{SYK})\frac{\pi \sqrt{2}}{(\beta\sqrt{\lambda}) ^{3/2}} e^{\frac{1}{\lambda}\left[2\beta\sqrt{\lambda} -\frac{\pi^2}{2}+\frac{\pi^2}{\beta  \sqrt{\lambda}}\right]},
\end{equation}
in the regime 
\begin{equation}\label{eqn:ncftRegime}
\lambda := -\log q \to 0^+,  \quad \lambda^{3/2} \ll \beta^{-1} \ll \lambda^{1/2}.
\end{equation}
This is also the result for the double-scaled SYK (DS-SYK), and hence the subscript on the left-hand side of \eqref{eqn:dssykParitionFunc}. 
Note in our normalization both the  temperature and the Hamiltonian are dimensionless.   The regime \eqref{eqn:ncftRegime} is what is called the NCFT$_1$ limit.  
It was also explained in \cite{Berkooz:2018qkz,berkooz2023parisis,berkooz2023parisislong} how to construct probe operators in such highly-fluxed models, that is, one essentially chooses operators of a similar  form as $H$ but with another flux $\tilde F_{IK}$ which may or may not correlate with $F_{IK}$. Moments with operator insertions take a slightly generalized form of equation \eqref{eqn:chordSchematic}.  It was demonstrated all such constructions have the same $n$-point correlation functions in the NCFT$_1$ limit \cite{Berkooz:2018qkz,Berkooz:2018jqr},  which are the same as what can be obtained from a universal Schwarzian effective action.   In particular, the correlation functions are conformal at leading order  in $\lambda$; at subleading order,   the out-of-time-ordered (OTO) four-point function has the special form
\begin{equation}
\vev{O(t)O(0)O(t)O(0)} \propto \lambda^{\#} \exp\left[\frac{2\pi}{\beta} t\right],
\end{equation}
which saturates the chaos bound of \cite{maldacena2015} and is a signature of the existence of a black hole horizon.

We wish to stress that what we just articulated are highly nontrivial requirements that would exclude a lot of models. For example,  pure random matrix models would strongly violate the uniformity condition on fluxes since hoppings carry an independently random phase on each Fock-space graph link, and local models strongly violate the condition of having a large amount of fluxes since the vast majority of terms in the Hamiltonian commute. The latter condition is also strongly violated by ``bosonic SYK'' type of models where canonical bosons are randomly coupled in a nonlocal manner similar to that of the usual fermionic SYK.   Indeed, although bosonic SYK models possess the same formal solutions as the fermionic ones,  these solutions are unphysical at low temperature due to the appearance of orderings (such as spin glass ordering) \cite{Georges2000,Georges2001, baldwin2020}.
Let us repeat a caveat mentioned in \cite{berkooz2023parisis,berkooz2023parisislong}, which is that the precise statements of the conditions are sufficient but not necessary for  NAdS$_2$/NCFT$_1$ dynamics. For example, though our conditions will include the double-scaled limit of the SYK model (to be introduced in section \ref{sec:examples}), it does not include  the usual large $N$ limit of a four-body SYK model.   The usual large $N$ limit of a four-body SYK model is a NCFT$_1$ model as well, but it would have $q=1$ which means it has less flux than we required. Nevertheless the four-body SYK supports a lot more random fluxes  in its Fock space than local models, and thus is a  weak violation of the precise conditions we laid out. So it seems to be just touching the edge (from the outside)  of our conditions. It is an open question whether we can find the sufficient and necessary conditions .

One may wonder if there is a dynamical mechanism for such fluxes to arise in a parent model of the NCFT$_1$, for example $\mathcal{N}=4$ super-Yang-Mills theory. It was conjectured  in \cite{berkooz2023parisis,berkooz2023parisislong} that adiabaticity and Berry phase may be such a mechanism. In the current paper, we will be satisfied by just treating it as a convenient tool for building NCFT$_1$ models and we will stay agnostic about its dynamical origin.
\subsection{Some known examples and the new models we will study}\label{sec:examples}
Let us mention a few known models that fit the general description we laid out in the last section. First there is the double-scaled limit of various $p$-local models, that is, models that randomly couple operators on $p$ different qubits in a $N$-qubit system with $p\ll N$. Double scaling means we scale $p$ as $p\sim \sqrt{N}$. For example, the SYK model is of the form of 
\begin{equation}
    H_{\text{SYK}} = i^{p/2}\sum_{1\leq i_1 <\ldots<i_p \leq N} J_{i_1\ldots i_p}\chi_{i_1} \cdots \chi_{i_p} 
\end{equation}
with even $p$, where $J_{i_1\ldots i_p}$ are independent real Gaussian random numbers with zero mean and $\chi$ are Majorana fermions. Its Fock space is generated by fermionic creation operators 
\begin{equation}
    \psi_i^\dagger = \chi_{2i-1} + i \chi_{2i}, \quad, i=1,\ldots, \floor{N/2}.
\end{equation}
The Hamiltonian is of the form of equation \eqref{eqn:generalHami} with 
\begin{equation}
    M_I = i^{p/2} J_{i_1\ldots i_p}\chi_{i_1} \cdots \chi_{i_p}, \quad I=\{i_1, \ldots,i_p\}.
\end{equation}
This satisfies 
\begin{equation}
    M_I M_K = (-1)^{|I\cap K|} M_K M_I =e^{i \pi |I\cap K|}M_K M_I, \quad M_I^2 \propto \mathbb{1}.
\end{equation}
which means $F_{IK} = \pi |I\cap K|$ (for such fluxes antisymmetry or symmetry in $IK$ make no difference).    If we choose the elements of $I$ and $K$ uniformly randomly from $\{1,\ldots,N\}$, it induces a probability distribution on $|I\cap K|$. If we further take the double-scaled limit 
\begin{equation}
    \text{fixed }  \frac{p^2}{N}, \ N\to \infty, 
\end{equation}
$|I\cap K|$ becomes an i.i.d. random variable such that \cite{erdos2014,cotler2016,garcia2017,garcia2018c,Berkooz:2018qkz}
\begin{equation}\label{eqn:qfactorSYK}
    q = \vev{\cos F_{IK} } = \binom{N}{p}^{-2}\sum_{I,K} (-1)^{|I\cap K|}  \to  e^{-2p^2/N} < 1.
\end{equation}
Hence the i.i.d. randomness and large amount condition on fluxes \eqref{eqn:randomFluxCond} is satisfied as well.  Here the random distribution on fluxes is induced from treating $I$ as a random set. Other double-scaled $p$-local models work in a very similar manner, such as where the basic variables are non-commuting Pauli matrices \cite{erdos2014, Berkooz:2018qkz}. Note if one decides to use $p$-local models to build up the Fock-space fluxes, the basic operators must have some degree of non-commutativity with phases, and hence pure canonical bosons will not work in the  $p$-local approach.

A second way of constructing Fock-space fluxes is in a sense more direct: we can place a lot of  fluxes  on a given Fock-space graph by assigning $U(1)$ link variables to the graph edges, and then directly assign an \textit{a priori}  probability distribution of the fluxes which is  i.i.d.. This would make $p$-locality or double scaling unnecessary. For example we can use the following fluxed operators as building blocks:
\begin{equation}
    T_i^+ = c_i^\dagger e^{\frac{i}{2} \sum_{k, k\neq i}^N F_{ik} r_k },  \quad T_i^- := (T_i^+)^\dagger.
\end{equation}
where $\{c_i^\dagger\}$ are the creation operators that generate the Fock space, and $\{r_k\}$ are the  site occupation number operators introduced in equation \eqref{eqn:localChargeOp}, whose sum is the total charge/number operator $R$.  
They satisfy
 \begin{equation}\label{eqn:fluxedhoppingAlg}
       T_i^\pm  T_j^\pm =  T_j^\pm  T_i^\pm e^{i F_{ij}}, \quad    T_i^\pm  T_j^\mp =  T_j^\mp  T_i^\pm e^{-i F_{ij}}, 
\end{equation}
regardless of the nature of $c_i^\dagger$ operators.  They can be derived by using equation \eqref{eqn:chargeOpFiniteTransform}. For example they can be fermions $\psi^\dagger_i$, bosons $b_i^\dagger$ or Pauli matrices $\sigma_i^+ := (\sigma_i^1 + i \sigma_i^2)/2$ (fermions bring an extra minus sign on the right-hand sides of \eqref{eqn:fluxedhoppingAlg}).   In all these cases (up to normal ordering and possibly reinterpreting a subset of $c^\dagger_k$ as anti-particle annihilation)
\begin{equation}
    r_k = c_k^\dagger c_k
\end{equation}
and 
\begin{equation}
T_k^+ T_k^- =  c_k^\dagger c_k.
\end{equation}
Clearly the resulting $T_i^\pm$ operators are not local at all,  specifically, they involve all the available $N$ sites together (so they are not $p$-local either).  They cannot be represented by some simple local operators by a Jordan-Wigner transformation since the phases are random.
The case when $c_i^\dagger =\sigma_i^+$ can be used to construct a known model of Parisi \cite{Parisi:1994jg,berkooz2023parisis,berkooz2023parisislong}:
\begin{equation}\label{eqn:parisiHami}
    H_\text{Parisi} =  -\frac{1}{\sqrt{N}} \sum_{i=1}^N (T_i^+ + T_i^-) 
\end{equation}
whose Fock-space graph is a simple $N$-dimensional hypercube.  
The $ T_i^\pm$ carry unit charge under $R$, i.e.  $[R, T_i^\pm] = \pm T_i^\pm$.  We can build charge-conserving models of the form
\begin{equation}\label{eqn:generalHamiPlocal}
    H = \sum_I M_I + h.c., \quad I=\{i_1,\ldots,i_{2p}\}, \quad M_I\propto  T_{i_1}^+\cdots  T_{i_p}^+ T_{i_{p+1}}^- \cdots T_{i_{2p}}^-.
\end{equation}
This is the type of models we will consider in this paper.  We will keep $p$ to be any order-one constant positive integer, and we emphasis again we do \textbf{not} need double scaling to reach a chord diagram description in this type of construction.  However double scaling is not forbidden either, and by this we mean two things: 1. the conventional double-scaled constructions such as the DSSYK is an example of Fock space fluxes as demonstrated at the beginning of this section.  2.  we can also let $p\sim \sqrt{N}$ in the construction of equation \eqref{eqn:generalHamiPlocal}. The moments will still take the general form of \eqref{eqn:chordSchematic},  but the relation between $q$ and $F_{ij}$ will be modified. Moreover, the thermodynamic analysis will follow more closely to that of \cite{Berkooz:2020uly} rather than the one we will see in the current paper.  And one of the points of the current paper is that fixing $p$ gives better-behaving thermodynamics. 

We have not specified what kind of index set $I$ to use.  We wish to consider simple models where the $M_I$'s pairwise contract at leading order in $1/N$.\footnote{Those that are not dominated by  pairwise contractions may still give  interesting (but different) nearly conformal physics.}  This does not impose a strong constraint on the possible structures of $I$ and indeed there are a great many choices. For example we can take $I$ to be of a ``chain'' form: 
\begin{equation}\label{eqn:Tchains}
    I = \{ i+1,  i+2 \ldots,i+2p \}, \quad i \in \{0, \ldots, N-2p\},
\end{equation}
which gives the Hamiltonian 
\begin{equation}\label{eqn:Hami2}
H = \frac{1}{\sqrt{2N}} \sum_{i=0}^{N-2p}  T_{i+1}^+T_{i+2}^+  \cdots T_{i+p}^+ T_{i+p+1}^-T_{i+p+2}^-\cdots T_{i+2p}^-+ h.c..
\end{equation}
The simplest case with $p=1$ gives
\begin{equation}\label{eqn:Hami2p=1}
H = \frac{1}{\sqrt{2N}} \sum_{i=0}^{N-2} \left( T_{i+1}^+T_{i+2}^-+  T_{i+2}^+T_{i+1}^-\right).
\end{equation}
If we wish we can even arrange the index set so that its elements form a local region on a $n$-dimensional lattice.   However let us remind ourselves that $T^\pm_i$ operators are already highly nonlocal, so we never end up with a truly local model, unless all the fluxes are set to zero.
Another way to achieve pairwise contractions is to use extra disorders, for example we can take
\begin{equation} \label{eqn:Hami1}
 H  = \sum_{1\leq i_1<\ldots < i_{2p}\leq N}J_{i_1 \ldots i_{2p}} T_{i_1}^+\cdots  T_{i_p}^+ T_{i_{p+1}}^- \cdots T_{i_{2p}}^- + h.c.
\end{equation}
where $J_{i_1 \ldots i_{2p}}$ are i.i.d. random numbers with zero mean. In this case there is no constraint on the choice of the index set $I$. 
Again, since $T^\pm_i$ operators are already highly nonlocal,  the choice \eqref{eqn:Hami1} is not $p$-local, unless we set all the fluxes to zero.

An interesting application of this construction is when we use bosonic creation operators $b_i^\dagger$ to construct the Fock space and the corresponding fluxed operators:
\begin{equation}
    [b_i, b_j] =0,  \quad [b_i, b_j^\dagger] =\delta_{ij}, \quad    T_i^+ = b_i^\dagger e^{\frac{i}{2} \sum_{k, k\neq i} F_{ik} b_k^\dagger b_k }.
\end{equation}
The NCFT$_1$ physics is guaranteed by the fluxes, and contrary to fermion/qubit-based models the Hilbert space dimension is infinite even at finite system size. Therefore we are allowed to take the particle density $\mathcal{Q}$ to be large which is impossible in fermion/qubit-based models.  
Even though our model is built from bosons, the large amount of fluxes in the Hilbert space severely frustrate the return amplitude of a wavefunction and hence cause delocalization, and this suggests that there is no unwanted low-temperature  ordering.  This is in  contrast with the bosonic SYK type of constructions.
  We  will make this argument sharp for the $p=1$ model defined in equation \eqref{eqn:Hami2p=1}.

For comparison purpose, we will also study a second class of charge-conserving models where $T_i^\pm$ are built from qubits. These models have similar charge scalings as the complex SYK.  It  enjoys some simplifications in the combinatorics compared to double-scaled complex SYK model and leads to a more regular behaviour in the free energy and the Lyapunov exponent, essentially because we do not need double scaling and can keep $p$ as an order-one constant.  As mentioned before,  the Fock space flux approach does not forbid double scaling, but if we were to double scale our construction,  such improvement will be lost.

\section{Models based on canonical bosons}\label{sec:bosonicModel}
In this section we discuss  a class of models built from canonical bosons.  In section \ref{sec:canBosonDef} we lay out the definitions,  and in section \ref{sec:canBosonPartiFunc} we  discuss in some detail on how to solve their partition functions  using  chord diagram  techniques. In section \ref{sec:canBosonNoOrdering} we will then demonstrate that they do not develop unwanted orderings at low temperatures.  In section \ref{sec:canBosonCorrelation}
we discuss how to use chord diagrams to solve for the correlation functions.  The gist is that all the moments will have a universal piece that is identical to the DS-SYK moments, and  from the DS-SYK moments one can obtain the correlation functions through the transfer matrix method \cite{Berkooz:2018jqr,Berkooz:2018qkz}.\footnote{Alternatively, one can use a chord path integral method \cite{berkooz2024chaos,berkooz2024path}.}
Finally in section \ref{sec:canBosonRMT}, we will compute the level statistics numerically at very low energies and show that their short-range correlations follow random matrix universality, which further corroborates the lack of ordering.  

As mentioned in section \ref{sec:examples}, the form of the Hamiltonian can be quite flexible as long as the basic building blocks are the fluxed operators $T_i^\pm$, and they all give the same leading order solutions. For presentation purpose we will use the chain form Hamiltonian \eqref{eqn:Hami2}.  One reason is that we want to maintain the visual distinction from $p$-local constructions: although the construction \eqref{eqn:Hami1} is not $p$-local either, it can be confused as one for a casual reading. The second reason is that although none of them develops low-temperature orderings, the argument is particularly sharp for the chain form Hamiltonian. 
\subsection{Definitions}\label{sec:canBosonDef}
Let us first  study the bosonic models mentioned in section \ref{sec:examples}. For clarity let us repeat the core definitions. We start with a collection of $N$ bosonic creation and annihilation operators with commutation relations
\begin{equation}
    [b_i,b_j]=0, \quad [b_i,b_j^\dagger] = \delta_{ij}.
\end{equation}
The fluxed operators in corresponding Fock space are 
\begin{equation}\label{eqn:bosonTopDef}
    T^+_i = b_i^\dagger e^{\frac{i}{2}\sum_{k, k \neq i} F_{ik} b_k^\dagger b_k},  \quad   T^-_i = b_i e^{-\frac{i}{2}\sum_{k, k \neq i} F_{ik} b_k^\dagger b_k}, 
\end{equation}
where $F_{ij}$ are antisymmetric in $i,j$ and are i.i.d. distributed for distinct pairs of $[ij]$.  We further assume the distribution is even in $F$ so that  $\langle\sin F_{ij} \rangle  =0$ and $\langle\cos F_{ij}\rangle $ is a tunable parameter.  The fluxed operators satisfy the algebra \eqref{eqn:fluxedhoppingAlg}, which we repeat here:
\begin{equation}\label{eqn:bosonFluxAlg}
    T^\pm_i   T^\pm_j =  e^{iF_{ij}} T^\pm_j   T^\pm_i, \quad T^\pm_i   T^\mp_j =  e^{-iF_{ij}} T^\mp_j  T^\pm_i, \quad i\neq j.
\end{equation}
Moreover
\begin{equation}
         T^+_i   T^-_i = b^\dagger_i b_i, \quad     T^-_i   T^+_i = b_i b_i^\dagger, 
\end{equation}
and 
\begin{equation}
     [b^\dagger_i b_i, T^\pm_j] = [b_i b_i^\dagger, T^\pm_j] = 0 \quad \text{for $i\neq j$}.
\end{equation}
The number operator is  
\begin{equation}
    R = \sum_{i=1}^N b_i^\dagger b_i
\end{equation}
whose eigenvalues are all the non-negative integers.   We will study the chain form Hamiltonian
\begin{align}
    H = \frac{1}{\sqrt{2N}} \sum_{i=0}^{N-2p}  T_{i+1}^+T_{i+2}^+  \cdots T_{i+p}^+ T_{i+p+1}^-T_{i+p+2}^-\cdots T_{i+2p}^-+ h.c. \nn 
\end{align}
which has already appeared as equation \eqref{eqn:Hami2}. We can use a short-hand notation 
\begin{equation}
   H = \frac{1}{\sqrt{2N}}  \sum_{I} (T_I + T^\dagger_I),
\end{equation}
where 
\begin{equation}
    I := \{i+1,i+2,\ldots, i+2p\}, \quad T_I :=  T_{i+1}^+  \cdots T_{i+p}^+ T_{i+p+1}^-\cdots T_{i+2p}^-, \quad  \sum_I := \sum_{i=0}^{N-2p}.
\end{equation}
As mentioned, many other choices have the same leading-order solution.\footnote{For the choice of \eqref{eqn:Hami1}, the equivalent normalization would be
\[ \vev{J_I^2} = \frac{1}{2}\binom{N}{2p}^{-1}.
\]
}
The energy operator in the grand canonical ensemble is 
\begin{equation}
H - \mu R,
\end{equation}
where $\mu$ is the chemical potential.  As we shall see the energy of $H - \mu R$ is bounded from below if $\mu$ is  negative and is order-one in $N$, even though the energy of $H$ per se is not bounded from below. Moreover, in the canonical ensemble (fixed total charge),  the energy of $H$ in this charge sector is bounded because every fixed charge sector is finite dimensional. 

\subsection{Moments, chords and thermodynamics}\label{sec:canBosonPartiFunc}
\begin{figure}
    \centering
    \includegraphics[scale=0.5]{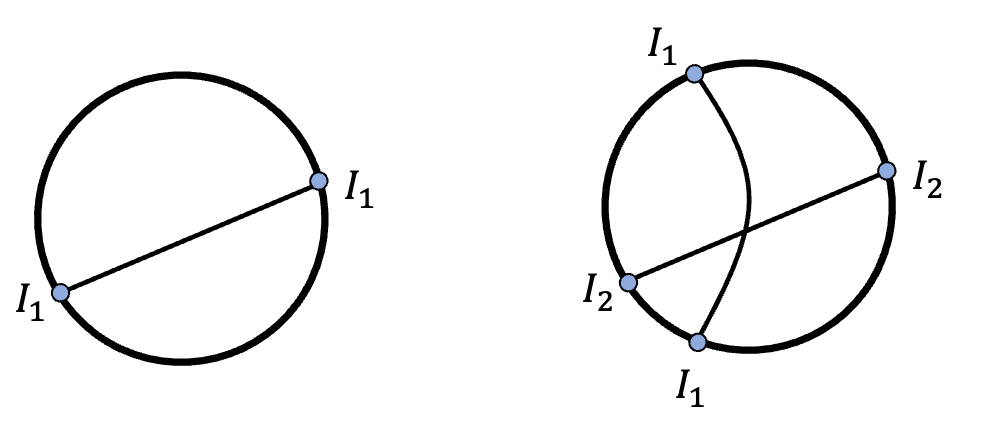}
    \caption{Two examples of chord diagrams.  Left: the diagram that gives $\vev{\Tr e^{aR }H^2} = \frac{1}{2N}\sum_{I_1}\langle\Tr e^{aR }(T_{I_1}+T_{I_1}^\dagger)^2\rangle$. Right: one of the three diagrams that contributes to $\vev{\Tr e^{aR }H^4}$,  this one represents  $\frac{1}{(2N)^2}\sum_{I_1,I_2}\langle\Tr e^{aR }(T_{I_1}+T_{I_1}^\dagger)(T_{I_2}+T_{I_2}^\dagger)(T_{I_1}+T_{I_1}^\dagger)(T_{I_2}+T_{I_2}^\dagger)\rangle$}
    \label{fig:Hchordexamples}
\end{figure}

The grand parition function is given by 
\begin{equation}
Z =  \vev{\Tr e^{a R - \beta H} }, 
\end{equation}
where $a=\mu \beta$.  To compute it let us consider the  average moments 
\begin{equation}\label{eqn:momDef1}
   m_k(a) := \vev{\Tr e^{a R} H^k}, \quad a<0,
\end{equation}
where the ensemble averaging is over all the fluxes $F_{ij}$. The $e^{aR}$ factor also plays the role of a regulator since $\Tr H^k$ itself is ill-defined.
To evaluate the  moments, we first note that  each $b_i$ must be paired with a $b_i^\dagger$ to give a nonzero trace, namely each index $i \in \{1,\ldots N\}$ must appear even number (including zero) of times.  For the chain form Hamiltonian this implies odd moments vanish.  
 For even moments $T_I+T_I^\dagger$ operators appear in pairs (Wick contractions), namely 
\begin{equation}\label{eqn:momWick}
m_k(a) =\frac{1}{(2N)^{k/2}} \sum_{\text{contractions}} \sum_{I_1,\ldots,I_{k/2}}\vev{\Tr\left[e^{a R} (T_{I_1}+T_{I_1}^\dagger) \ldots (T_{I_{k/2}}+T_{I_{k/2}}^\dagger)\ldots \right]}, 
\end{equation}
where $k$ is even, and there are $k$ factors of $T_I+T_I^\dagger$ whose subscripts form $k/2$ pairs.  There are $(k-1)!!$ number of such contractions. Each contraction can be represented diagrammatically, namely, we represent the trace by a circle, and draw points on the circle which represent the subscripts of the $T_I$'s.  Then we connect two points from inside the circle by a line,  whenever the subscripts they represent are paired.  Diagrams obtained this way are called \textit{chord diagrams}, and in figure \ref{fig:Hchordexamples} we show two examples of such diagrams.
 We are interested in the large $N$ limit of above expression for all $k$,  and if we ever need to take  $k\to \infty$  we only do so after taking the large $N$ limit.  In this sense we always have $k= O(1)$ in $N$.  In such a regime,  at leading order in $1/N$ we can assume every two sets $I_i$ and $I_j$ in $\{I_1, \ldots, I_{k/2}\}$ to have empty intersection\footnote{An index $i$ can appear more than two times (say to form a term like $(b_i b_i^\dagger)^2$),  but the sum of all such terms is $1/N$ suppressed by simple counting.} 
\begin{equation}\label{eqn:emptyInters}
|I_i \cap I_j| =0.
\end{equation}
 From the basic relations \eqref{eqn:bosonFluxAlg}, we can derive the following algebra among $T_I$'s:
\begin{equation}\label{eqn:fluxedAlgMultiIndex}
\begin{split}
     T_I T_K &= \exp\left[i\sum_{l=1}^p \sum_{m=1}^{p}\left(F_{i_l k_m}+ F_{i_{l+p} k_{m+p}} -F_{i_{l+p} k_m}-F_{i_l k_{m+p}} \right)\right]T_K T_I, \\
     T_I T_K^\dagger &= \exp\left[-i\sum_{l=1}^p \sum_{m=1}^{p}\left(F_{i_l k_m}+ F_{i_{l+p} k_{m+p}} -F_{i_{l+p} k_m}-F_{i_l k_{m+p}} \right)\right] T_K^\dagger T_I,
\end{split}
\end{equation}
for $|I \cap K| =0$. We have used the notation 
\begin{equation}
    i_l := i+l, \quad k_m := k+m
\end{equation}
and so on.  Hence the effective fluxes are 
\begin{equation}
F_{IK} = \sum_{l=1}^p  \sum_{m=1}^{p}\left(F_{i_l k_m}+ F_{i_{l+p} k_{m+p}} -F_{i_{l+p} k_m}-F_{i_l k_{m+p}}\right).
\end{equation}
This in turn gives 
\begin{align}\label{eqn:adjacentProduct}
    &(T_I + T_I^\dagger)  ( T_K + T_K^\dagger) 
    = ( T_K + T_K^\dagger) (T_I + T_I^\dagger)\cos F_{IK}   +i  ( T_K - T_K^\dagger) (T_I - T_I^\dagger)   \sin F_{IK}.
\end{align}
Since  $F_{ij}$ are assumed to have an i.i.d. even distribution,  this gives us an i.i.d.  even distribution on all the $F_{IK}$ (with $|I\cap K|=0$) which has
\begin{equation}
q:=\vev{\cos F_{IK}}=\vev{\cos F_{ij}}^{4p^2}, \quad  \vev{ \sin F_{IK}} = 0.
\end{equation}
We can move the $T_I + T_I^\dagger$ factors across each other until every contracted pairs are adjacent to each other, and in the process we gain phase factors according to equation \eqref{eqn:adjacentProduct}.
We have argued that at leading order in $1/N$,  every  two subscript sets among $I_1,\ldots, I_{k/2}$ in equation \eqref{eqn:momWick} have empty intersection. This in particular implies $(\cos F_{ij})^2$ or  $(\sin F_{ij})^2$ (or higher powers) never appear at the leading order.\footnote{One does not need to  worry about, say,  passing a $T_I$ through  two $T_K$'s  because $T_K^2$ traces to zero inside moments.  What can happen is passing a $T_I$ through a $T_K$ and then a $T_K^\dagger$, but in this case the two acquired phases cancel according to  equation \eqref{eqn:fluxedAlgMultiIndex}.} Hence we can safely ignore all the sine terms since they ensemble-average to zero, and all the cosine terms that appear will have distinct pairs of subscripts on the fluxes and hence factorize under averaging because of the i.i.d. assumption.  Since each intersection in a chord diagram  represents an interlacing ordering of the subscripts on the product of $T$'s,  each intersection would give rise to a factor of $q = \langle \cos F_{IK} \rangle$.  Thus we have 
\begin{equation}\label{eqn:contractionqPower}
\text{a contraction } \propto  q^\text{number of intersections in the corresponding chord diagram}.
\end{equation}
The proportionality constant of the above expression is
\begin{equation}\label{eqn:momPropfactor}
\frac{1}{(2N)^{k/2}} \sum_{I_1,\ldots, I_{k/2}}\Tr\left( e^{aR} \prod_{m=1}^{k/2} (T_{I_m}+ T_{I_m}^\dagger)^2\right).
\end{equation}
Now 
\begin{equation}\label{eqn:bsquares}
\begin{split}
(T_{I}+ T_{I}^\dagger)^2 = &T_{I}^2  + [ (b_{i_1}^\dagger b_{i_1}) \cdots (b_{i_{p}}^\dagger b_{i_{p}} )]  [(b_{i_{p+1}} b_{i_{p+1}}^\dagger)\cdots  (b_{i_{2p}} b_{i_{2p}}^\dagger)] \\
&+ [ (b_{i_1}b_{i_1}^\dagger ) \cdots ( b_{i_{p}} b_{i_{p}}^\dagger)]  [( b_{i_{p+1}}^\dagger b_{i_{p+1}})\cdots  ( b_{i_{2p}}^\dagger b_{i_{2p}})]+ (T_{I}^\dagger)^2.
\end{split}
\end{equation}
The $T_I^2$ and its hermitian conjugate cannot contribute to the trace.
The bosonic Fock space is a tensor product of single-particle space $\mathcal{H}_i, i=1,\ldots N$,  and hence everything boils down to three types of traces
\begin{equation}\label{eqn:threeTracesBoson}
\begin{split}
    \Tr_{\mathcal{H}_i}(e^{a b_i^\dagger b_i}) & = \frac{1}{1-e^{a}},\\
        \Tr_{\mathcal{H}_i}(e^{a b_i^\dagger b_i}b_i^\dagger b_i) & = \frac{e^{a}}{(1-e^{a})^2},\\
          \Tr_{\mathcal{H}_i}(e^{a b_i^\dagger b_i} b_ib_i^\dagger) & = \frac{1}{(1-e^{a})^2}.
\end{split}
\end{equation}
Combining equation \eqref{eqn:contractionqPower}--\eqref{eqn:threeTracesBoson},  we get the final result for even moments (odd moments vanish) at leading order in $1/N$,
\begin{equation}\label{eqn:momWithChemPotential}
\begin{split}
       m_k(a)& =\left[ \frac{1}{1-e^{a}}\right]^{N- p k} \left[\frac{e^{a}}{(1-e^{a})^4}\right]^{kp/2}\sum_{CD_{k}} q^{\text{No. of int.}} \\
       & = \left[ \frac{1}{1-e^{a}}\right]^{N} \left[\frac{e^{a}}{(1-e^{a})^2}\right]^{kp/2}\sum_{CD_{k}} q^{\text{No. of  int.}}.
       \end{split}
\end{equation}
where $\sum_{CD_{k}}$ means a sum over chord diagrams with $k$ points on the circle (or $k/2$ chords), and  ``No. of int.'' means the number of chord intersections. We reiterate that all the above derivation holds for  the Hamiltonian \eqref{eqn:Hami1} as well.
Note that the expression $\sum_{CD_{k/2}}q^{\text{No. of  int.}}$   exactly  computes the moments of the double-scaled SYK (DS-SYK)  introduced in section \ref{sec:examples}:
\begin{align}
m_k^{\text{DSSYK}}:=& \vev{\Tr H_\text{SYK}^k} \qquad \qquad\qquad \text{($q=e^{-2p^2/N}$ fixed,  $N \to \infty$)} \nn \\
 =&\text{dim} (\mathcal{H}_\text{SYK})  \sum_{CD_{k}}q^{\text{No. of int.}}.
\end{align}
Therefore,  we have the relation 
\begin{equation}\label{eqn:momRelationChem}
     m_k(a) =\left[ \frac{1}{1-e^{a}}\right]^{N} \left[\frac{e^{a}}{(1-e^{a})^2}\right]^{kp/2} \text{dim} (\mathcal{H}_\text{SYK})^{-1}  m_k^{\text{DSSYK}}.
\end{equation}
The partition function is a Taylor series with moments being the coeffients:
\begin{equation}
Z(\beta) = \sum_k m_k \frac{(-\beta)^k}{k!}.
\end{equation}
Given relation \eqref{eqn:momRelationChem}, this implies
\begin{equation}
Z(\beta, a) = \left[ \frac{1}{1-e^{a}}\right]^{N}   \text{dim} (\mathcal{H}_\text{SYK})^{-1}  Z_\text{DSSYK}\left(\beta \left[\frac{e^{a}}{(1-e^{a})^2}\right]^{p/2}\right),
\end{equation}
where  the NCFT$_1$ limit of the DS-SYK partition function is given in equation \eqref{eqn:dssykParitionFunc}.  Again, setting $a=\mu\beta$ gives the partition function of $H$ at chemical potential $\mu$.

The moment computation is similar but much simpler than that of the  double-scaled charged SYK model (DS-cSYK) presented in \cite{Berkooz:2020uly}, and the result is free from several peculiarities of \cite{Berkooz:2020uly}. In the double-scaled construction,  a nontrivial $q$ factor is obtained by summing the holonomies over all possible values of $|I \cap K|$ with a Poisson weight (recall equation \eqref{eqn:qfactorSYK}).  With a chemical potential, each different $|I \cap K|$ brings about a different factor dependent on  the chemical potential.  The upshot is that \cite{Berkooz:2020uly} ends up with a $q$ factor that is dependent on the chemical potential:
\begin{equation}
    q_\text{DS-cSYK}(a) = e^{-\frac{4p^2}{N} (\cosh a)^2}.
\end{equation}
Moreover, the $k$-th moment picks up an extra factor of
\begin{equation}
    (e^{\frac{p^2}{2N} (\sinh a)^2})^{k^2}
\end{equation}
from the non-intersecting pairs of chords.  This causes several issues: first, it is hard to deal with a order-one chemical potential because the partition function is no longer easily summable in $k$ due to the $k^2$ exponent. Second,  if we go to the canonical ensemble, the $q$-factor develops a charge dependence which makes it impossible to take a NCFT$_1$ limit for the sectors with near-maximal/minimal charges. Third, the fact that $p$ goes to infinity makes the quantities such as ground state energy and Lyapunov exponent in fixed charge sectors develop singular behaviours.  Our construction is free of all these problems because our $p$ is a fixed order-one integer, and the leading moments only involve contributions from $|I \cap K| =0$.  We have already seen our grand partition function has a simple form for any given chemical potential,  now let us go to the canonical ensemble with fixed charges. 
Let us compute the thermodynamics in a sector with a fixed particle number $Q$,  whose density is nonzero, i.e.,
\begin{equation}\label{eqn:chargeRange}
    Q =N\mathcal{Q}, \quad \mathcal{Q}\in (0,+\infty).
\end{equation}
Note the $\mathcal{Q}\to \infty$ limit is impossible in qubit- or fermion-based models because the latter have finite-dimensional Hilbert spaces at finite $N$.  Hence equation \eqref{eqn:chargeRange} is useful in itself because we can now engineer broader scaling regimes.

The easiest route to the physics of fixed $\mathcal{Q}$ sectors is to consider moments in each sector:
\begin{align}\label{eqn:fixedchargemomIntegral1}
  m_k(\mathcal{Q}) &:= \vev{\delta(R-N\mathcal{Q}) H^k} = \frac{1}{2\pi} \int_{-\pi}^\pi da e^{-iN\mathcal{Q} a} m_k( ia)   \\
             &=   \text{dim} (\mathcal{H}_\text{SYK})^{-1} m_k^{\text{DSSYK}} \times \frac{1}{2\pi} \int da\left[\frac{e^{ia}}{(1-e^{ia})^2}\right]^{kp/2}   \exp\left[-iN\mathcal{Q} a -N\log(1-e^{ia})\right], \nn
\end{align}
where the integration range is $[-\pi, \pi)$ because $Q$ is an integer.  We can then do a saddle-point evaluation of the integral  at large $N$.   A caveat here is that the formal integral in the second line of \eqref{eqn:fixedchargemomIntegral1} is divergent due to the singularity at $a=0$.  This may seem a little strange,  since $R= Q$ enforces that we are tracing over a finite-dimensional subspace and the result must be finite.
 However if we are clear-eyed about what we are actually doing,  we will see doing saddle-point evaluation is fine even though the formal integral presented is ill-defined. We explain why this is the case in appendix \ref{app:formalIntegral}.  Now we shall simply proceed with the saddle point analysis.
 Note that $k$ is order-one in $N$ (as explained in the paragraph above equation \eqref{eqn:emptyInters}),  so it does not enter the saddle-point equation.   The saddle point is given by
\begin{equation}\label{eqn:bosonChemSaddle}
   a_{\text{saddle}} = i \log(1+ \mathcal{Q}^{-1}).
\end{equation}
The moments evaluate to 
\begin{equation}
     m_k(\mathcal{Q}) = e^{N\mathcal{Q}\log(1+ \mathcal{Q}^{-1})+ N \log(1+\mathcal{Q})} [\mathcal{Q}(1+\mathcal{Q})]^{pk/2}   \text{dim} (\mathcal{H}_\text{SYK})^{-1} m_k^{\text{DSSYK}}.
\end{equation} 
This means 
\begin{equation}\label{eqn:bosonpartiRelation1}
Z(\beta, \mathcal{Q}) = e^{N\mathcal{Q}\log(1+ \mathcal{Q}^{-1})+ N \log(1+\mathcal{Q})}   \text{dim} (\mathcal{H}_\text{SYK})^{-1}  Z_\text{DSSYK}\left(\beta [\mathcal{Q}(1+\mathcal{Q})]^{p/2}\right).
\end{equation}
In the NCFT$_1$ limit \eqref{eqn:ncftRegime},  we can obtain the extremal entropy and extremal energy by combining equation \eqref{eqn:bosonpartiRelation1} and equation \eqref{eqn:dssykParitionFunc}:
\begin{equation}\label{eqn:extremalScalingBoson1}
    S_0(\mathcal{Q}) = N\mathcal{Q}\log(1+ \mathcal{Q}^{-1})+ N \log(1+\mathcal{Q}) - \frac{\pi^2}{2\lambda}, \quad  E_0(\mathcal{Q}) =- \frac{2}{\sqrt{\lambda}} [\mathcal{Q}(1+\mathcal{Q})]^{p/2},
\end{equation}
where $\lambda :=\log q \to 0^+$.   
 Note we have always taken the $N\to \infty$ limit first,  and the $\lambda\to 0$ limit after.  This means 
$N \gg \lambda^{-1}$.
Hence the  dominant parts of the extremal entropy are just the terms proportional to $N$.  Unfortunately the scaling behaviour in $\mathcal{Q}$  does not match with that of black holes.
Since the Fock space is bosonic, the density $\mathcal{Q}$ can be arbitrarily large.   Note since our $p$ is an order-one integer,  $E_0$ is a well-defined quantity. In the double-scaled (or large $p$) complex SYK, the ground state energies of all charge sectors collapse to zero as $p\to \infty$ \cite{Berkooz:2020uly, davison2017}. For example in the double-scaled complex SYK, we have the formal expression \cite{Berkooz:2020uly}
\begin{equation}
     E^\text{DScSYK}_0(\mathcal{Q}) =-\frac{2}{\sqrt{\lambda}}(1-4\mathcal{Q}^2)^\frac{p+1}{2},  \quad |\mathcal{Q}|\leq \frac{1}{2}, \quad p\to \infty.
\end{equation}
Our case is much better behaved since $p$ is finite.  This is not an issue for finite-$p$ complex SYK either, but there the drawback is we do not have a simple analytic expression for $E_0$ \cite{davison2017,gu2020}.
Our expression of $E_0(\mathcal{Q})$  shows that even though the total spectrum of $H$ is not bounded from below, it is for a fixed $\mathcal{Q}$, so the canonical ensemble is well defined. 
In the grand canonical ensemble,   the Hamiltonian is shifted to $H- \mu R$.  This means the ground state energy gets shifted to $-\mu N\mathcal{Q} +E_0$, and since $N \gg \lambda^{-1}$, the shifted energy is bounded below if $\mu$ is negative and order-one, therefore the grand canonical ensemble is well-defined too.  This reasoning is  based on the annealed-averaged calculation that gives $E_0$. In the next section we argue that there is no ordering,  which implies that the annealed result can be trusted.  In the special case of $p=1$ we give a rigorous proof for the lower-boundedness of $H-\mu R$.
\subsection{No low-temperature ordering or energetic instability}\label{sec:canBosonNoOrdering}
We argue that there is no low-temperature ordering for our model.  The gist is that the very large amount random fluxes in the Fock space significantly frustrate the return amplitude of a many-body wavefunction and thus cause delocalization. Constructive interferences are avoided by virtue of the fluxes being i.i.d. random. If there is no ordering, the equation \eqref{eqn:extremalScalingBoson1} for  the annealed many-body ground state energy $E_0$ with fixed $\mathcal{Q}$ should be trusted,  and hence there are no energetic instabilities if we shift it to $E_0-\mu N \mathcal{Q}$ by a negative chemical potential.
We can give a rigorous proof for the lower-boundedness of $H-\mu R$, and a particularly sharp argument for the lack of ordering,   for the special case where $p=1$:
\begin{equation}\label{eqn:p=1chain}
H = \frac{1}{\sqrt{2N}} \sum_{i=0}^{N-2}  T_{i+1}^+ T_{i+2}^- + h.c.
\end{equation}
If we turn off all the fluxes,  the Hamiltonian $H$ becomes 
\begin{equation}\label{eqn:p=1chainNoFlux}
    H_0 =  \frac{1}{\sqrt{2N}} \sum_{i=0}^{N-2} b_{i+1}^\dagger b_{i+2} + h.c. 
\end{equation}
This is just a nearest-neighbor hopping model on a chain,  which can be put into a free boson gas form 
\begin{equation}
 \sum_{i} \varepsilon_i  \tilde b_i^\dagger \tilde b_i
\end{equation}
where $\varepsilon_i$ are single-particle energies,   and $\tilde b_i$ are related to $b_i$ by an orthogonal transformation of the subscript and $\tilde b_i$  still satisfy the canonical commutation relations. 
With a chemical potential, the energy operator at zero flux  then has the form
\begin{equation}
   H_0  - \mu R =  \sum_i (\varepsilon_i -\mu ) b_i^\dagger b_i.
\end{equation}
The smallest single-particle energy of $H_0$ behaves as
\begin{equation}
\min(\{\varepsilon_i\}) \sim - \frac{1}{\sqrt{N}}.
\end{equation}
Hence $H_0-\mu R$ is bounded from below  since $\mu$ is an order-one negative constant. In fact, this  shows $\mu$ can be taken to be as small as $\sim 1/\sqrt{N}$ to stabilize the Hamiltonian. Next we show that the ground state energy of $H-\mu R$ cannot be smaller than that of $H_0- \mu R$,  which is  analogous to diamagnetism.\footnote{The mathematics is exactly the same. However here the fluxes are interpreted to be in the Hilbert space not real space,  so there is no actual magnetic field involved.}  Note for both $H$ and $H_0$  we have
\begin{align}
\Tr(e^{\mu \beta R } H^{2k+1}) =\Tr(e^{\mu \beta R} H_0^{2k+1}) =0
\end{align} 
even without disorder averaging and at finite $N$.
For even moments we have a diamagnetic inequality
\begin{equation}
\Tr(e^{\mu \beta R } H^{2k}) \leq \Tr(e^{\mu \beta R} H_0^{2k}),
\end{equation}
essentially because whenever there is a $e^{i F}$ term in the expression of the left-hand side as a sum over loops, 
there is a corresponding term on  the right-hand side which simply evaluates to 1.    We then exponentiate this inequality to get
\begin{equation}
\Tr(e^{-\beta (H-\mu R) } ) \leq \Tr(e^{-\beta (H_0-\mu R) }).
\end{equation}
Taking the $\beta \to \infty$ limit establishes our claim.  Regarding ordering,  since $H_0 -\mu R$ describes a free boson gas in one dimension,   it is already free of ordering.   Now adding lots of random fluxes in the Fock space only makes the many-body wavefunctions more delocalized  as explained at the beginning of this section, which should make it even harder for ordering to happen. Thus, in the fluxed model we expect no ordering. None of the above relies on ensemble averaging and thus holds for each single realization of the system. Note that given our normalization,  the lack of ordering holds true for temperatures as low as $\sim 1/\sqrt{N}$. Namely, it should not be understood as a trivial effect of a would-be ordering temperature getting suppressed to zero by normalization, which sometimes does happen for spin glass models  in the double scaling limit (e.g., see appendix A of \cite{berkooz2024path}).  We will corroborate this claim in section \ref{sec:canBosonRMT} by studying the energy spacing ratio statistics, which is independent of the overall normalization of the Hamiltonian.

The above proof should be adaptable to the Hamiltonians of the type \eqref{eqn:Hami2} for any order-one value of $p$ because the zero-flux case is also a one-dimensional translational invariant chain, which should not support any ordering by Hoehenberg-Mermin-Wigner type of considerations. 
Note we did not use the full strength of the assumption that the large amount of random Fock-space fluxes prohibit orderings. Rather we picked some special representatives of the permissble Hamiltonians,  such that when fluxes are turned off the Hamiltonians are free of disorder  and become exactly solvable. The zero-flux solution exhibits an energetic lower bound and a lack of ordering.  The properties of the random fluxes that were actually used in the proof were diamagnetism and that they should not increase the chance of ordering.  For the Hamiltonians of the type  \eqref{eqn:Hami1} this approach may not work since the zero-flux case is a $p$-local bosonic-SYK type  model, which is known to have orderings \cite{Georges2000,Georges2001}. Nevertheless we believe the conclusion shall remain the same once the fluxes are turned on, though a proof may become harder. This will be supported by the numerical evidences in section \ref{sec:canBosonRMT}.

\subsection{Correlation functions}\label{sec:canBosonCorrelation}
We follow \cite{berkooz2023parisis,berkooz2023parisislong} for the definition of probe operators.  That is,  probes are operators built from 
\begin{equation}\label{eqn:bosonProbTopDef}
 \tilde  T^+_i = b_i^\dagger e^{\frac{i}{2}\sum_{k, k \neq i} \tilde F_{ik} b_k^\dagger b_k},  \quad  \tilde  T^-_i = b_i e^{-\frac{i}{2}\sum_{k, k \neq i} \tilde F_{ik} b_k^\dagger b_k}.
   \end{equation}
   They are of the same form as $T_i^\pm$ operators defined in equation \eqref{eqn:bosonTopDef}, only with a second flux $\tilde{F}_{ik}$.  We again require $\tilde{F}_{ik}$ to be i.i.d.  distributed, and the distribution is even in $\tilde F_{ik}$.  However,  we do not require  $\tilde{F}_{ik}$ to be statistically independent from  ${F}_{ik}$.    The $ \tilde  T^\pm_i$ satisfy the same algebra among themselves as those  among $T^\pm_i$, that is, of equation \eqref{eqn:bosonFluxAlg} with $F$ replaced by $\tilde{F}$.  In addition they satisfy the following relations with $T_i^\pm$:
   \begin{equation}\label{eqn:mixedFluxAlg}
 T^\pm_i \tilde  T^\pm_j = e^{i \frac{F_{ij}+ \tilde F_{ij}}{2}}  \tilde  T^\pm_j  T^\pm_i , \quad  T^\pm_i \tilde  T^\mp_j = e^{-i \frac{F_{ij}+ \tilde F_{ij}}{2}}  \tilde  T^\mp_j  T^\pm_i.
   \end{equation}
We may consider probe operators of the form 
   \begin{equation}\label{eqn:bosonProbesDef}
 O_{p_+,p_-}:= \frac{1}{\sqrt{N}}\sum_{K_{p_+, p_-}}  \tilde{T}_{K_{p_+,p_-}} ,
   \end{equation}
   where 
   \begin{equation}
   \tilde{T}_{K_{p_+, p_-}}:=\tilde T_{k+1}^+ \cdots \tilde T_{k+{p_+}}^+ \tilde T_{k+ p_++1}^- \cdots \tilde T_{k+p_+ + p_-}^-,\, \quad \sum_{K_{p_+, p_-}} = \sum_{k=0}^{N-p_+-p_-}.
   \end{equation}
  The probe $O_{p_+,p_-}$ is a sum over monomials of $p_+ + p_-$ basic fluxed operators and carries a charge of $p_+ - p_-$ , and the choice  is by no means   exhaustive.    The motivation for such a choice is that we think of both the Hamiltonian and the probes as the infrared operators descending from some single-trace operators in a UV holographic CFT (Hamitonian descends from the stress-energy tensor), with a highly excited near-extremal state as the background.  Therefore they should belong to a class of statistical operators that look similar as the infrared Hamiltonian, and hence the choice \cite{Berkooz:2018qkz,Berkooz:2018jqr}.  The one-point (or any odd-point) moment of $\tilde T$ is exponentially suppressed (schematically of the form $\langle \cos(F-\tilde F)\rangle^N$ \cite{berkooz2023parisis,berkooz2023parisislong}), so we can just focus on the even-point functions.
\subsubsection{Two-point functions and spectral asymmetry}  
\begin{figure}
    \centering
    \includegraphics[scale=0.5]{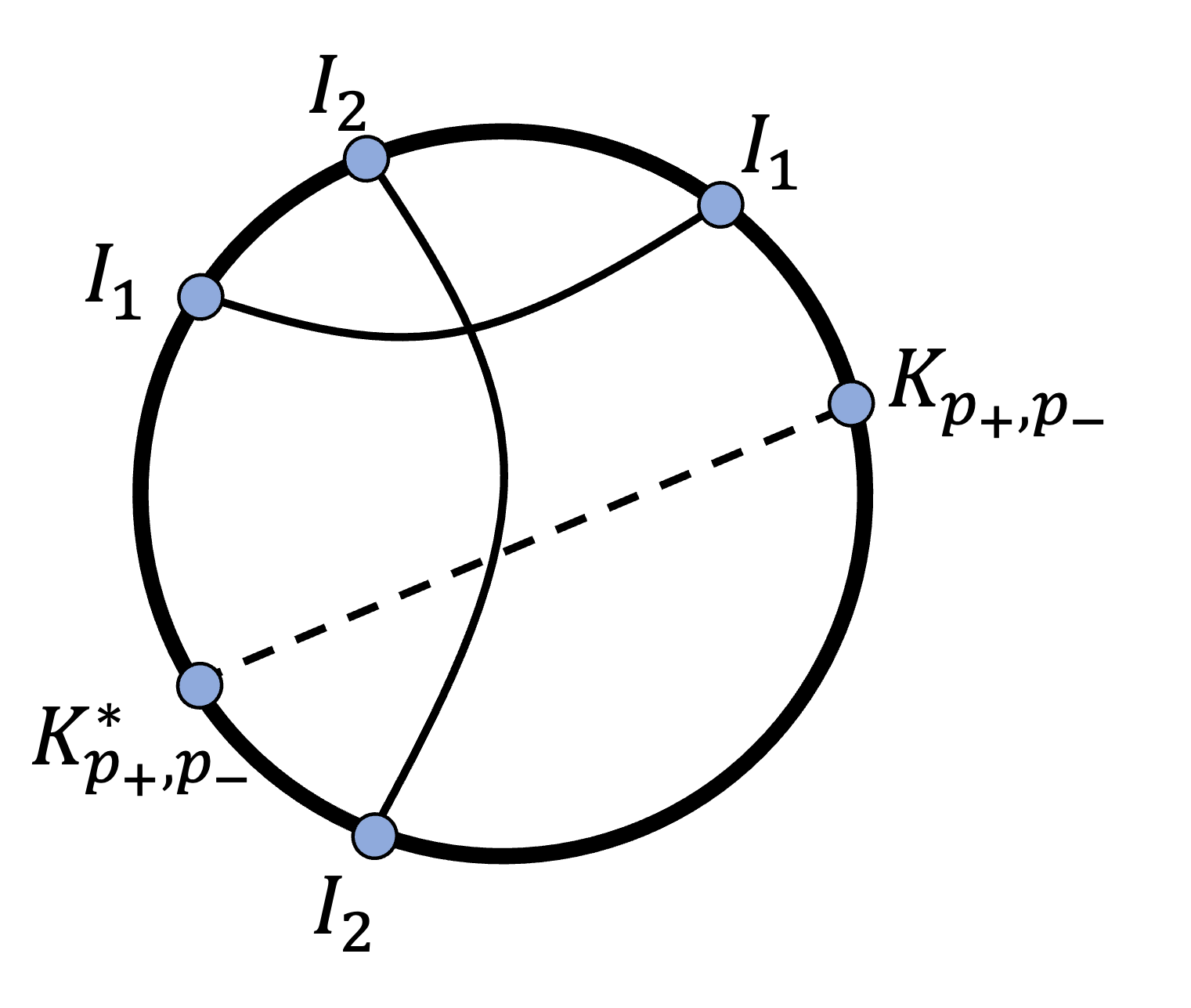}
    \caption{Chord diagram representation of a two-point insertion.  The solid chords are from Hamiltonian insertions and the dashed chord is from probe insertions.  This diagram in particular represents $\sum_{I_1,I_2,K_{p_+,p_-}}\langle\Tr e^{aR }(T_{I_1}+T_{I_1}^\dagger)(T_{I_2}+T_{I_2}^\dagger)(T_{I_1}+T_{I_1}^\dagger) \tilde T_{K_{p_+,p_-}}(T_{I_2}+T_{I_2}^\dagger)\tilde T_{K_{p_+,p_-}}^\dagger\rangle/(4N^3)$, which contributes to $\langle \Tr e^{aR} H^3 O_{p_+,p_-}HO_{p_+,p_-}^\dagger \rangle$, i.e., $m_{3,1}^{p_+,p_-}$.}
    \label{fig:twopoint}
\end{figure}

 The  time-ordered Euclidean two-point functions at finite chemical potential defined as
  \begin{equation}
 G(\tau,\beta,\mu)= \frac{1}{Z(\beta,\mu)}\vev{\Tr(e^{-\beta (H-\mu R)}\mathcal{T}_\tau \left[e^{(H-\mu R) \tau }   O_{p_+,p_-} e^{-(H-\mu R) \tau }   O_{p_+,p_-}^\dagger )\right]}.
  \end{equation}
Specifically
  \begin{equation}
  G(\tau>0,\beta,\mu)=  \frac{e^{(p_+-p_- )\mu\tau}}{Z(\beta,\mu)}\vev{\Tr(e^{-\beta (H-\mu R)} e^{H \tau }   O_{p_+,p_-} e^{-H \tau }   O_{p_+,p_-}^\dagger )},
  \end{equation}
  and
   \begin{equation}
  G(\tau<0,\beta,\mu)=  \frac{e^{-(p_+-p_- )\mu\tau}}{Z(\beta,\mu)}\vev{\Tr(e^{-\beta (H-\mu R)}   O_{p_+,p_-}^\dagger e^{H \tau }   O_{p_+,p_-} e^{-H \tau }  )},
  \end{equation}
  We can get the  two-point functions by studying their moments
   \begin{equation}
   m_{k_1,k_2}^{p_+,p_-}(a) := \vev {\Tr e^{aR}  H^{k_1} O_{p_+, p_-} H^{k_2} O_{p_+, p_-}^\dagger }
 \end{equation}
and 
 \begin{equation}
   m_{k_1,k_2}^{p_-,p_+}(a) := \vev {\Tr e^{aR}  H^{k_1} O^\dagger _{p_+, p_-} H^{k_2} O_{p_+, p_-}}.
 \end{equation}
   From the basic algebra \eqref{eqn:mixedFluxAlg} for two types of fluxes, we can derive 
  \begin{equation}
  T_I \tilde{T}_{K_{p_+, p_-}} = \tilde{T}_{K_{p_+, p_-}}   T_I \exp\left[i\left(\sum_{n=1}^{p_+}\sum_{m=1}^p -\sum_{n=1}^{p_+}\sum_{m=p+1}^{2p}+ \sum_{n=p_++1}^{p_++p_-}\sum_{m=p+1}^{2p} -\sum_{n=p_++1}^{p_++p_-}\sum_{m=1}^{p} \right)\frac{F_{i_m k_n}+\tilde F_{i_m k_n}}{2}\right].
  \end{equation}
  Again we have used the notation $i_m := i+m, k_n := k+n$ and so on.
  The average of the phase factor on the right-hand side is 
  \begin{equation}
  \tilde q = \vev{\cos \frac{F_{ik}+\tilde F_{ik}}{2}}^{2p(p_+ + p_-)}.
  \end{equation}
Playing the same game as we did for deriving chord rules \eqref{eqn:momWithChemPotential},  we get the following for two-point insertions:
   \begin{equation}\label{eqn:bosonTwoPtMomChem}
   m_{k_1,k_2}^{p_+,p_-}(a)=\left[ \frac{1}{1-e^{a}}\right]^{N} \left[\frac{e^a}{(1-e^{a})^2}\right]^{\frac{p(k_1+k_2)}{2}}
  \left[\frac{1}{1-e^{a}}\right]^ {p_++p_-} e^{ap_+}  \text{dim} (\mathcal{H}_\text{SYK})^{-1}  m_{k_1,k_2}^{\text{DSSYK}},
 \end{equation}
and
 \begin{equation}\label{eqn:bosonTwoPtMomChem2}
   m_{k_1,k_2}^{p_-,p_+}(a)=e^{a(p_--p_+)}  m_{k_1,k_2}^{p_+,p_-}(a),
 \end{equation}
 where 
 \begin{equation}
  \text{dim} (\mathcal{H}_\text{SYK})^{-1}  m_{k_1,k_2}^{\text{DSSYK}} = \sum_{CD_{k_1,k_2}}  q^{\text{No. of $H$-$H$ int}}\ \tilde q^{\text{No. of $H$-$O$ int}}.
 \end{equation}
 Here $CD_{k_1,k_2}$ means all chord diagrams with one $O$-chord and $(k_1+k_2)/2$ $H$-chords,  with the $O$-chord splitting the end points of $H$-chords into $k_1$ points and $k_2$ points.  We give a example of such in the figure \ref{fig:twopoint}.    
  For fixed charge sectors (canonical ensembles), the two-point function is defined as
 \begin{equation}
 G(\tau,\beta,\mathcal{Q})_{p_+,p_-}= \frac{1}{Z(\beta,\mathcal{Q})}\vev{\Tr(\delta(R-N\mathcal{Q})e^{-\beta H}\mathcal{T}_\tau \left[e^{H \tau }   O_{p_+,p_-} e^{-	H \tau }   O_{p_+,p_-}^\dagger )\right]}.
  \end{equation}
 We can get to fixed charge sectors by Fourier-transforming equation \eqref{eqn:bosonTwoPtMomChem}.    
 The saddle point for $a$ remain the same as  equation \eqref{eqn:bosonChemSaddle}.  Hence 
    \begin{equation}\label{eqn:bosonTwoPtMomCharge}
   m_{k_1,k_2}^{p_+,p_-}(\mathcal{Q})=e^{S_0(\mathcal{Q})} [\mathcal{Q}(1+\mathcal{Q})]^{p(k_1+k_2)/2}  (1+\mathcal{Q})^{p_++p_-} (1+\mathcal{Q}^{-1})^{p_+} \text{dim} (\mathcal{H}_\text{SYK})^{-1} m_{k_1,k_2}^{\text{DSSYK}},
 \end{equation}
and 
     \begin{equation}\label{eqn:bosonTwoPtMomChargeRetard}
   m_{k_1,k_2}^{p_-,p_+}(\mathcal{Q})=e^{S_0(\mathcal{Q})} [\mathcal{Q}(1+\mathcal{Q})]^{p(k_1+k_2)/2}  (1+\mathcal{Q})^{p_++p_-} (1+\mathcal{Q}^{-1})^{p_-} \text{dim} (\mathcal{H}_\text{SYK})^{-1} m_{k_1,k_2}^{\text{DSSYK}}.
 \end{equation}
Hence we have the following relations in the fixed-charge sectors:
\begin{equation}
\begin{split}
G(\tau>0,\beta, \mathcal{Q})_{p_+,p_-} &=  (1+\mathcal{Q})^{p_++p_-} (1+\mathcal{Q}^{-1})^{p_+}  G_\text{DSSYK}([\mathcal{Q}(1+\mathcal{Q})]^{\frac{p}{2}}\tau,[\mathcal{Q}(1+\mathcal{Q})]^{\frac{p}{2}}\beta),\\
  G(\tau<0,\beta, \mathcal{Q})_{p_+,p_-} &=  (1+\mathcal{Q})^{p_++p_-} (1+\mathcal{Q}^{-1})^{p_-}  G_\text{DSSYK}([\mathcal{Q}(1+\mathcal{Q})]^{\frac{p}{2}}\tau,[\mathcal{Q}(1+\mathcal{Q})]^{\frac{p}{2}}\beta).
\end{split}
\end{equation}
where $ G_\text{DSSYK}$ is the two-point function for DS-SYK.  In the NCFT$_1$ regime 
\begin{equation}
q, \tilde q \to 1^-,   (-\log q)^{-1/2} \ll \beta \ll (-\log q)^{-3/2},
\end{equation}
this implies that the two-point functions has the same conformal form as the Majorana DS-SYK model, aside from a $\mathcal{Q}$-dependent wavefunction normalization. In particular, the conformal dimension of a probe is
 \begin{equation}\label{eqn:confDimBoson}
     \Delta_{O_{p_+,p_-}} = \frac{\log \tilde q}{\log q} = \frac{p_+ + p_-}{8p}\frac{\langle (F_{ik}+\tilde F_{ik})^2\rangle}{\langle F_{ik}^2\rangle}.
 \end{equation}
 The difference between the advanced and the retarded pieces is charaterized by the ratio
 \begin{equation}
 \frac{ G(\tau>0,\beta, \mathcal{Q})_{p_+,p_-}}{ G(-\tau<0,\beta, \mathcal{Q})_{p_+,p_-} } =  e^{(p_+ - p_-)\log (1+ \mathcal{Q}^{-1})}.
 \end{equation}
This gives the spectral asymmetry parameter $\mathcal{E}$ \cite{faulkner2011, sachdev2015prx}
 \begin{equation}\label{eqn:bosonSpecAsymm}
       e^{2\pi (p_+-p_-)\mathcal{E}} =  e^{(p_+ - p_-)\log (1+ \mathcal{Q}^{-1})}.
\end{equation}
This spectral asymmetry satisfies the relation
 \begin{equation}
 \mathcal{E} = \frac{1}{2\pi} \frac{dS_0(\mathcal{Q})/N}{d\mathcal{Q}}.
 \end{equation}
 \subsubsection{Four-point functions and maximal chaos}\label{sec:maximalChaosBoson}
 We are intereseted in two types of four-point functions: uncrossed and crossed.   Uncrossed four-point functions are of the form 
\begin{equation}
    \vev{O_1(\tau_4) O^\dagger_1(\tau_3) O_2(\tau_2) O^\dagger_2(\tau_1)}_\beta \quad \text{or} \quad    \vev{O_1(\tau_4) O_2(\tau_3) O_2^\dagger(\tau_2) O_1^\dagger(\tau_1)}_\beta
\end{equation}
and crossed four-point functions are of the form 
\begin{equation}
      \vev{O_1(\tau_4) O_2(\tau_3)  O^\dagger_1(\tau_2) O^\dagger_2(\tau_1)}_\beta.
\end{equation}
Again, we define the probes $O_i$ as we did in equation \eqref{eqn:bosonProbesDef} with fluxes $\tilde F^{(i)}$, and carry $p_{i+}$ number of $\tilde T^+$ and $p_{i-}$ number of $\tilde T^-$. 
They require the computation of (uncrossed and crossed) four-point moments. The computation is entirely similar to that of the two-point moments. At a given chemical potential, we have 
\begin{align}\label{eqn:bosonUncrossed4ptMom}
   m_{k_1,k_2,k_3,k_4}^\text{uncrossed} (a) = &\vev{\Tr e^{a R}H^{k_1}O_1H^{k_2}O_1^\dagger H^{k_3}O_2H^{k_4} O_2^\dagger } \\
= &\left[ \frac{1}{1-e^{a}}\right]^{N} \left[\frac{e^a}{(1-e^{a})^2}\right]^{\frac{p}{2} \sum_{i=1}^4 k_i}
  \left[\frac{1}{1-e^{a}}\right]^ {\sum_{j=1}^2 (p_{j+}+p_{j-})} e^{a\sum_{j=1}^2p_{j+}} \nn \\
& \quad \times \sum_{CD_{k_1,k_2,k_3,k_4}^\text{uncrossed}} q^{\text{No. of $H$-$H$ int.}}\tilde q_1^{\text{No. of $H$-$O_1$ int. }} \tilde q_2^{\text{No. of $H$-$O_2$ int.}}, \nn
\end{align}
where 
    \begin{equation}
\tilde q_i := \vev{\cos \frac{F_{ik}+\tilde F_{ik}^{(i)}}{2}}^{2p(p_{i+} +p_{i-})}.
\end{equation}
Simlilarly for the crossed moments, we have 
\begin{align}\label{eqn:bosonCrossed4ptMom}
   m_{k_1,k_2,k_3,k_4}^\text{crossed} (a) = &\vev{\Tr e^{a R}H^{k_1}O_1H^{k_2} O_2 H^{k_3}O_1^\dagger H^{k_4} O_2^\dagger } \\
= &\left[ \frac{1}{1-e^{a}}\right]^{N} \left[\frac{e^a}{(1-e^{a})^2}\right]^{\frac{p}{2} \sum_{i=1}^4 k_i}
  \left[\frac{1}{1-e^{a}}\right]^ {\sum_{j=1}^2 (p_{j+}+p_{j-})} e^{a\sum_{j=1}^2p_{j+}} \nn \\
& \quad \times \tilde q_{12} \sum_{CD_{k_1,k_2,k_3,k_4}^\text{crossed}} q^{\text{No. of $H$-$H$ int.}}\tilde q_1^{\text{No. of $H$-$O_1$ int. }} \tilde q_2^{\text{No. of $H$-$O_2$ int.}} , \nn
\end{align}
where the intersection between the $O_1$-chord and $O_2$-chord gives rise to a new $q$-parameter
  \begin{equation}
\tilde q_{12} := \vev{\cos \frac{\tilde F^{(1)}+\tilde F^{(2)}}{2}}^{(p_{1+} +p_{1-})(p_{2+} +p_{2-})}.
\end{equation}
In both cases, other than the overall $a$-dependent prefactor, they are identical with the corresponding uncrossed and crossed four-point moments of DS-SYK, so we can write
\begin{equation}
    m_{k_1,k_2,k_3,k_4}(\mu) =\left[ \frac{1}{1-e^{a}}\right]^{N} \left[\frac{e^a}{(1-e^{a})^2}\right]^{\frac{p}{2} \sum_{i=1}^4 k_i}
  \left[\frac{1}{1-e^{a}}\right]^ {\sum_{j=1}^2 (p_{j+}+p_{j-})} e^{a\sum_{j=1}^2p_{j+}}   m_{k_1,k_2,k_3,k_4}^{\text{DSSYK}}. 
\end{equation}
The four-point moments in a fixed charge sector is then
\begin{align}
        m_{k_1,k_2,k_3,k_4}(\mathcal{Q})=e^{S_0(\mathcal{Q})} [\mathcal{Q}(1+\mathcal{Q})]^{\frac{p}{2}\sum_{i=1}^4 k_i } & (1+\mathcal{Q})^{\sum_{j=1}^2 (p_{j+}+p_{j-})} (1+\mathcal{Q}^{-1})^{\sum_{j=1}^2 p_{j+}} \nn \\
        &\qquad \qquad \times \text{dim} (\mathcal{H}_\text{SYK})^{-1} m_{k_1,k_2,k_3,k_4}^{\text{DSSYK}}
\end{align}
Again, for the correlation functions  the entropic factor  will cancel with the same contribution in the denominator, and we end up with 
\begin{align}\label{eqn:correlationRelationBoson}
     G(\tau_i,\beta,\mathcal{Q}) = (1+\mathcal{Q})^{\sum_{j=1}^2 (p_{j+}+p_{j-})} (1+\mathcal{Q}^{-1})^{\sum_{j=1}^2 p_{j+}}  G_\text{DSSYK}([\mathcal{Q}(1+\mathcal{Q})]^{\frac{p}{2}}\tau_i,[\mathcal{Q}(1+\mathcal{Q})]^{\frac{p}{2}}\beta),
\end{align}
which holds for both crossed and uncrossed types.  
It is not hard to see how this relation is generalized to arbitrary $2n$-point functions: just replace $\sum_{j=1}^2 p_{j\pm} $ by $\sum_{j=1}^{n} p_{j\pm}$ in the prefactors.   Note we chose a particular ordering of probes where $O_i$ always appear to the left of $O_i^\dagger$.  If we want a reverse ordering, each pair would bring an extra spectral asymmetry factor computed in equation \eqref{eqn:bosonSpecAsymm}.

If we consider the  out-of-time-ordered connected four-point function in the conformal regime, since
\begin{equation}
        \vev{O_1(t_1) O_2(0) O_1(t_2)O_2(0)}_\text{DSSYK} \sim (-\log q)^{\text{const}}\exp\left[\frac{2\pi}{\beta} \left(\frac{t_1+t_2}{2}\right)\right]
\end{equation}
at early time in the NCFT$_1$ limit, the relation \eqref{eqn:correlationRelationBoson} implies that our model has the same maximal Lyapunov exponent  
\begin{equation}
    \lambda_L =  \lambda_L^\text{SYK} = \frac{2\pi}{\beta} + O(\beta^{-2}),
\end{equation}
independent of the charge density, and this independence holds for higher-order (in temperature) corrections to $\lambda_L$.  In the DS-cSYK model, this is not true \cite{Berkooz:2020uly}: 
\begin{equation}\label{eqn:lyapunovDScYK}
    \lambda_L^\text{DScSYK} = \frac{2\pi}{\beta} - \frac{4\pi}{\beta^2 J \sqrt{-\log q}}  (1-4\mathcal{Q}^2)^{(1-p)/2}+ O(\beta^{-3}),
\end{equation}
where $p$ is the number of fermions in the  Hamiltonian.  In the double scaled limit $p$ is infinite, so it is a little hard to interpret the correction term since it diverges.  Our model does not suffer from this, because the number of hoppings (in both the Hamiltonian and the probes) is kept at order one.

\subsection{Spectral statistics and random matrix universality}\label{sec:canBosonRMT}
Let us demonstrate that the short-range correlations of the energy levels follow the random matrix theory (RMT) universality.  This reflects the very long-time behaviour of our system, and ascribes to another notion of quantum chaos universality through the Bohigas-Giannoni-Schmit conjecture \cite{bohigas1984}. If we denote the energy eigenvalues as $E_i$, $i=1,2,\ldots$ in increasing order, namely  $E_i<E_{i+1}$,  then the nearest level spacing is
\begin{equation}
 \delta E_i: = E_{i+1} -E_i, 
\end{equation}
We define the  spacing ratio $r$ as the following
\begin{equation}
r_i : = \frac{\text{min}( \delta E_i, \delta E_{i+1})}{\text{max}( \delta E_i, \delta E_{i+1})},
\end{equation}
and we will study the statistics of such spacing ratios in a fixed-charge sector.   The advantage of using spacing ratio is that the dependence on the average spectral density is canceled out, so that we do not need to perform unfolding on the spectrum.  For Gaussian unitary ensemble (GUE) of random matrices,  an approximate analytic formula for the distribution of the spacing ratios is given by \cite{atas2016}
\begin{equation}\label{eqn:spacingRatioGUE}
\rho_\text{GUE}(r) =   \frac{C(r + r^2)^2 }{ (1 + r + r^2)^4},
\end{equation}
where $C$ is a normalization factor so that $\rho_\text{GUE}$ integrates to one.  Although the above expression is approximate,  the error turns out to be rather small,  which is  similar to the situation of Wigner's surmise as an approximation of RMT level spacing distribution.  We now compare $\rho_\text{GUE}(r)$ with the numerical spacing ratio distribution $\rho(r)$ obtained from the $p=1$ Hamiltonian \eqref{eqn:p=1chain}, which we used to argue for the absence of orderings at low temperatures: 
\[
H = \frac{1}{\sqrt{2N}} \sum_{i=0}^{N-2}  T_{i+1}^+ T_{i+2}^- + h.c.
\]
We work with $N=20$ and in the sector with occupation number $Q=6$, whose dimensionality is 177100, and we obtain the lowest 100 eigenvalues (the lowest 0.06\% of all levels) from 2000 realizations of the Hamiltonian.  We choose the fluxes $F_{ij}$ to follow a uniform distribution supported on $[-0.2\pi, 0.2\pi]$.\footnote{So $q=\langle \cos F_{ij}\rangle^4 = [\sin (0.2\pi)/ (0.2\pi)]^4 \approx 0.766$.}  In computing the spacing statistics, we also exclude the lowest 20 eigenvalues for the following reasons:
\begin{enumerate}
    \item  The near-CFT$_1$ regime \eqref{eqn:ncftRegime} does not include the extremely low-energy states.
    \item  The  formula \eqref{eqn:spacingRatioGUE} is not expected to hold in the first place for levels extremely close to the spectral edge even if a system is fully chaotic \cite{altland2024quantum}.\footnote{In our case, there are still level repulsions in the lowest 20 eigenvalues but the repulsions are much weaker.}
\end{enumerate}
\begin{figure}[t]
    \centering
    \includegraphics[width=\textwidth]{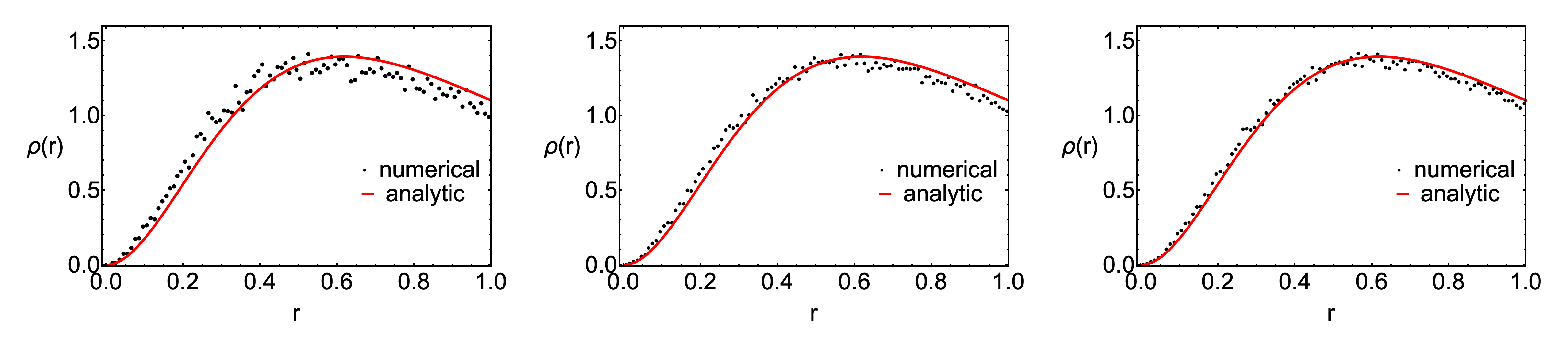}
    \caption{Low-energy spacing ratio statistics for 2000 realizations of the Hamiltonian \eqref{eqn:p=1chain} ($N=20, Q=6$). The black dots are the numerical results and the red curve is the GUE analytic result \eqref{eqn:spacingRatioGUE}.  Left: computed from the lowest 40 levels. Middle: computed from the lowest 70 levels. Right: computed from the lowest 100 levels.  All  have excluded the contribution from the lowest 20 levels. }
    \label{fig:canBosonRatios}
\vspace{1cm}
    \includegraphics[width=\textwidth]{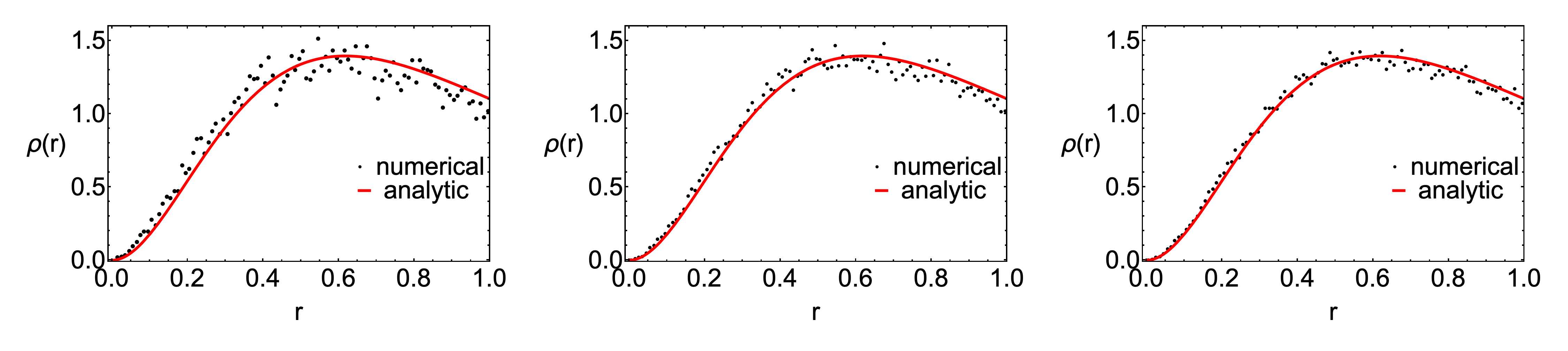}
     \caption{Low-energy spacing ratio statistics for 1500 realizations of the Hamiltonian with Gaussian couplings, $H= \sum_{i<j}^{N}J_{ij} T_{i}^+ T_{j}^- + h.c.$ ($N=20, Q=6$). The black dots are the numerical results and the red curve is the GUE analytic result \eqref{eqn:spacingRatioGUE}.  Left: computed from the lowest 40 levels. Middle: computed from the lowest 70 levels. Right: computed from the lowest 100 levels.  All  have excluded the contribution from the lowest 20 levels. }
    \label{fig:canBosonRatiosGau}
\end{figure}
In figure \ref{fig:canBosonRatios} we compare the numerical results of the spacing ratio statistics with the analytic formula  \eqref{eqn:spacingRatioGUE} from Gaussian unitary random matrices. We can see the agreement with RMT is good, and the convergence to the RMT result quickly improves as we include more levels (all with the lowest 20 levels excluded). Since the spectrum exhibits RMT universality,   we expect our model to eigenstate-thermalize,  down to the very low energies. This further corroborates that our bosonic model does not have low-temperature orderings.
We also obtained the lowest 100 eigenvalues of 1500 realizations of the fluxed Hamiltonians with Gaussian couplings $J_{ij}$,
\[
H =\sum_{i<j}^{N} J_{ij} T_{i}^+ T_{j}^- + h.c.,
\]
and we plot the spacing ratio statistics in figure \ref{fig:canBosonRatiosGau}. The results are essentially the same as those of the chain form Hamiltonian.  This supports the claim made by the end of section \ref{sec:canBosonNoOrdering} that the low-temperature dynamics is always dominated by the fluxes even when there are random couplings.

We have not discussed the long-range fluctuations of the spectrum, which is not visible through spacing ratio statistics. Spectral form factor and number variance are better diagnostic tools for such correlations.  We would expect long-range correlations to manifest as early-time bumps in the connected spectral form factor and as a large tail in the number variance  \cite{Jia_2020,Jia:2019orl}.
\section{Models based on qubits}\label{sec:qubitModel}
In qubit-based models, we start by constructing fluxed hoppings using Pauli matrices:
\begin{equation}
    T_i^+ = \sigma_i^+ e^{\frac{i}{4} \sum_{k, k\neq i}^N F_{ik} \sigma^3_k }, \quad  \sigma^+:= \frac{1}{2}(\sigma^1 + i \sigma^2),
\end{equation}
and $T_i^-$ is the hermitian conjugate of $T_i^+$.  This is the same basic hopping operators used in constructing the Parisi hypercube model \cite{berkooz2023parisis,berkooz2023parisislong} (also see equation \eqref{eqn:parisiHami}).   The conserved charge operator we shall use is
\begin{equation}
    R = \frac{1}{2}\sum_{i=1}^N \sigma^3_i,
\end{equation}
which has the basic property
\begin{equation}
    [R, T^\pm_j] = \pm T^\pm_j.
\end{equation}
The charge density $\mathcal{Q}$ are the eigenvalues of $R/N$, which take values in $[-1/2, 1/2]$.   In terms of the $T^\pm_i$ operators, we can consider the same construction of Hamiltonians as the bosonic case
\begin{equation}
H = \frac{2^p}{\sqrt{2N}}\sum_{i=0}^{N-2p}  T_{i+1}^+T_{i+2}^+  \cdots T_{i+p}^+ T_{i+p+1}^-T_{i+p+2}^-\cdots T_{i+2p}^-+ h.c.,
\end{equation}
where $p$ is an order-one integer.   Note here we take the normalization of the Hamiltonian to be such that 
\begin{equation}
   2^{-N }\vev{\Tr H^2} =1.
\end{equation}
at leading order in $1/N$.\footnote{For the Hamiltonian \eqref{eqn:Hami1} this amounts to taking
\begin{equation}
    \vev{J_I^2} = 2^{2p-1}\binom{N}{2p}^{-1},
\end{equation}
}
And as before, this is only a representative of a large class of constructions which all have the same chord combinatorics. 
\subsection{Thermodynamics and correlation functions}
The fluxed algebra remain the same as equation \eqref{eqn:fluxedAlgMultiIndex} and hence the chord diagram combinatorics and the $q$ factors are also the same.   The only difference here is the thermodynamic factors that are functions of chemical potentials (or charges).  In terms of chemical potential,  instead of equation \eqref{eqn:threeTracesBoson} we now have
\begin{equation}\label{eqn:threeTracesQubits}
\begin{split}
    \Tr_{\mathcal{H}_i}(e^{a \sigma_i^3/2}) & = 2\cosh(a/2),\\
        \Tr_{\mathcal{H}_i}(e^{a  \sigma_i^3/2}\sigma_i^+ \sigma_i^-) & = e^{a/2},\\
          \Tr_{\mathcal{H}_i}(e^{a \sigma_i^3/2} \sigma_i^- \sigma_i^+) & = e^{-a/2}.
\end{split}
\end{equation}
Thus here we have 
\begin{equation}
 m_k(a)=\vev {\Tr H^k e^{ a R}} =2^N[\cosh  (a/2)]^{N- pk} \sum_{CD_k} q^{ \text{No. of  int.}} = [\cosh  (a/2)]^{N- pk}m_k^{\text{DSSYK}},
\end{equation}
where here 
\begin{equation}
m_k^{\text{DSSYK}} = 2^N  \sum_{CD_k} q^{\text{No. of  int.}}.
\end{equation}
Therefore at finite chemical potential,  the parition function is related to the DS-SYK partition function by 
\begin{equation}
       Z(\beta,a = \beta\mu)=  (\cosh \beta \mu/2)^{N}Z^{\text{DSSYK}}\left(\frac{\beta J}{[\cosh (\beta \mu/2)]^p} \right).
\end{equation}
Similar to the boson case, we can study a fixed charge sector with charge $Q= N \mathcal{Q}$,
\begin{align}\label{eqn:fixedchargemomIntegral2}
  m_k(\mathcal{Q}) &=  \int_{-\pi}^\pi da e^{-i N \mathcal{Q} a} m_k( ia)    = m_k^{\text{DSSYK}} \times \int da   (\cos a/2)^{N-pk} e^{-i N \mathcal{Q} a}.
\end{align}
The saddle point is given by 
\begin{equation}
    a_\text{saddle}  = i \log \frac{1-2\mathcal{Q}}{1+2\mathcal{Q}}\quad  \implies \cos \frac{a_\text{saddle}}{2} = \frac{1}{\sqrt{1-4\mathcal{Q}^2}},
\end{equation}
and 
\begin{equation}
  m_k(\mathcal{Q}) =e^{N\mathcal{Q}\log \frac{1-2\mathcal{Q}}{1+\mathcal{Q}} +N\log \frac{1}{
  \sqrt{1-4\mathcal{Q}^2}}} \left( 1-4\mathcal{Q}^2\right)^{\frac{pk}{2}}  m_k^{\text{DSSYK}}.
\end{equation}
This implies 
\begin{equation}
    Z(\beta, \mathcal{Q}) =  e^{ N \mathcal{Q}\log \frac{1-2\mathcal{Q}}{1+2\mathcal{Q}} +N\frac{1}{2}\log \frac{1}{1-4\mathcal{Q}^2} \ }  Z^\text{DSSYK}(\beta(1-4\mathcal{Q}^2)^{\frac{p}{2}}).
\end{equation}
Again using equation \eqref{eqn:dssykParitionFunc}, we can work out  the  zero-temperature entropy and energy  in the NCFT$_1$ regime,
\begin{align}
    S_0(\mathcal{Q}) &= N\left(\mathcal{Q}\log \frac{1-2\mathcal{Q}}{1+2\mathcal{Q}} + \frac{1}{2}\log \frac{4}{1-4\mathcal{Q}^2}\right)- \frac{\pi^2}{2\lambda}, \label{eqn:zeroTEntropyFixedCharge} \\
     E_0(\mathcal{Q}) &=-\frac{2}{\sqrt{\lambda}}(1-4\mathcal{Q}^2)^\frac{p}{2} \label{eqn:zeroTEnergyFixedCharge}
\end{align}
where we have absorbed the $2^N$ overall normalization factor from $ Z^\text{DSSYK}$ into the  $\log [{4}/(1-4\mathcal{Q}^2)]$ term in $S_0$.    The above expressions are formally quite similar to the corresponding results for the complex DS-SYK model \cite{Berkooz:2020uly}:\footnote{Their definition of $\lambda$ has a factor of 4 difference than ours, which we changed to our definition here.}
\begin{equation}
\begin{split}
    S^\text{DScSYK}_0 &= N\left(\mathcal{Q}\log \frac{1-2\mathcal{Q}}{1+2\mathcal{Q}} + \frac{1}{2}\log \frac{4}{1-4\mathcal{Q}^2}\right)- \frac{\pi^2}{2\lambda}(1-4\mathcal{Q}^2),  \\
     E^\text{DScSYK}_0 &=-\frac{2}{\sqrt{\lambda}}(1-4\mathcal{Q}^2)^\frac{p+1}{2}.
\end{split}
\end{equation}
This is also the result for the large $p$ limit (after large $N$ limit is taken) of the complex SYK model \cite{davison2017}.   A crucial difference is in the ground state energy:  in the complex DS-SYK model,  $p$ is formally infinite.   Therefore  $E^\text{DScSYK}_0$ collapses to 0  for any nonzero $\mathcal{Q}$.    In our construction,   $p$ is an order-one integer and hence does not suffer from this singular behaviour.

The probe operators  $O_{p_+, p_-}$ are constructed in the same way as the ones constructed in section \ref{sec:canBosonCorrelation} for canonical bosons.    The computation of the correlation functions is a straightforward repetition as well,  with the bosonic thermodynamic prefactors replaced by the qubit ones.  Here we simply state the results.  For two-point insertions we have
 \begin{align}
    m_{k_1,k_2}^{p_+,p_-}(a) :=& \vev {\Tr e^{aR}  H^{k_1} O_{p_+, p_-} H^{k_2} O_{p_+, p_-}^\dagger }  \nn \\
    =&2^N \cosh(a/2)^{N-p(k_1+k_2)-p_+ - p_-} e^{a (p_+ - p_-)/2}\sum_{CD} q^{\text{No. of $H$-$H$ int.}} \tilde q^{\text{No. of $H$-$O$ int.}}  \nn\\
     =&\cosh(a/2)^{N-p(k_1+k_2)-(p_+ + p_-)} e^{a (p_+ - p_-)/2}m_{k_1,k_2}^{\text{DSSYK}} 
 \end{align}
and 
 \begin{align}
    m_{k_1,k_2}^{p_-,p_+}(a) :=& \vev {\Tr e^{aR}  H^{k_1} O^\dagger _{p_+, p_-} H^{k_2} O_{p_+, p_-}} \nn \\
    =&\cosh(a/2)^{N-p(k_1+k_2)-p_+ - p_-} e^{-a (p_+ - p_-)/2}m_{k_1,k_2}^{\text{DSSYK}}  \\
    =& e^{-a (p_+ - p_-)}  m_{k_1,k_2}^{p_+,p_-}(a). \nn
 \end{align}
In charge-$N\mathcal{Q}$ sectors, the two-point moments are
\begin{align}\label{eqn:twoPtMomCharge}
 m^{p_+,p_-}_{k_1,k_2}(\mathcal{Q})=    e^{\frac{1}{2 }(p_- - p_+)\log \frac{1-2\mathcal{Q}}{1+2\mathcal{Q}}}e^{ N [\mathcal{Q}\log \frac{1-2\mathcal{Q}}{1+2\mathcal{Q}} +\frac{1}{2}\log \frac{1}{1-4\mathcal{Q}^2}]}  (1-4\mathcal{Q}^2)^{\frac{p(k_1+k_2)+p_+ +p_-}{2}} m^\text{DSSYK}_{k_1,k_2},
\end{align}
and 
\begin{equation}
 m^{p_-,p_+}_{k_1,k_2}(\mathcal{Q}) =  e^{(p_+ - p_-) \log \frac{1-2\mathcal{Q}}{1+2\mathcal{Q}}} m^{p_+,p_-}_{k_1,k_2}(\mathcal{Q}).
\end{equation}
Hence the two-point functions obey
\begin{equation}
\begin{split}
G(\tau>0,\beta, \mathcal{Q})_{p_+,p_-} &= e^{\frac{1}{2 }(p_- - p_+)\log \frac{1-2\mathcal{Q}}{1+2\mathcal{Q}}}  (1-4\mathcal{Q}^2)^{\frac{p_+ +p_-}{2}}  G_\text{DSSYK}((1-4\mathcal{Q}^2)^{\frac{p}{2}}\tau,(1-4\mathcal{Q}^2)^{\frac{p}{2}}\beta),\\
  G(\tau<0,\beta, \mathcal{Q})_{p_+,p_-} &=e^{\frac{1}{2 }(p_+ - p_-)\log \frac{1-2\mathcal{Q}}{1+2\mathcal{Q}}}  (1-4\mathcal{Q}^2)^{\frac{p_+ +p_-}{2}}  G_\text{DSSYK}((1-4\mathcal{Q}^2)^{\frac{p}{2}}\tau,(1-4\mathcal{Q}^2)^{\frac{p}{2}}\beta).
\end{split}
\end{equation}
This implies the conformal dimensions $\Delta_{p_+,o_-}$ have the same expression as equation \eqref{eqn:confDimBoson}.  Moreover,  now the asymmetry between the advanced and retarded correlations is 
 \begin{equation}
 \frac{ G(\tau>0,\beta, \mathcal{Q})_{p_+,p_-}}{ G(-\tau<0,\beta, \mathcal{Q})_{p_+,p_-} } =  e^{(p_+ - p_-)\log \frac{1-2\mathcal{Q}}{1+2\mathcal{Q}}},
 \end{equation}
from which we can read off the spectral asymmetry parameter:
\begin{equation}
\mathcal{E} = \frac{1}{2\pi}\log \frac{1-2\mathcal{Q}}{1+2\mathcal{Q}}.
\end{equation}
This  again satisfies 
 \begin{equation}
 \mathcal{E} = \frac{1}{2\pi} \frac{dS_0(\mathcal{Q})/N}{d\mathcal{Q}}.
 \end{equation}
 The computation quite easily extends to $2n$-point functions.  If we place all the $O_{p_{i+},p_{i-}}$ to the left of all the  $O_{p_{i+},p_{i-}}^\dagger$,  we will have 
 \begin{align}\label{eqn:correlationRelationQubit}
     G(\tau_i,\beta,\mathcal{Q}) = e^{\frac{1}{2 }\sum_{i=1}^n (p_{i-} - p_{i+})\log \frac{1-2\mathcal{Q}}{1+2\mathcal{Q}}}  (1-4\mathcal{Q}^2)^{\frac{1}{2 }\sum_{i=1}^n (p_{i-} + p_{i+})}  G_\text{DSSYK}((1-4\mathcal{Q}^2)^{\frac{p}{2}}\tau_i,(1-4\mathcal{Q}^2)^{\frac{p}{2}}\beta).
\end{align}
If we want to switch the ordering of a pair of $O_{p_{i+},p_{i-}}$ and  $O_{p_{i+},p_{i-}}^\dagger$, we simple multiply the above expression by the  $e^{2\pi \mathcal{E}(p_{i+}-p_{i-})}$.

Regarding four-point functions and Lyapunov exponent,  we have the same comments as the ones written at the end of section \ref{sec:maximalChaosBoson}. Namely, here we have the maximal Lyapunov exponent with a completely regular subleading-in-temperature correction, because our $p$ is an order-one number.  This is in sharp contrast with the DS-cSYK model \cite{Berkooz:2020uly}. 
\subsection{Spectral statistics}
\begin{figure}[t!]
     \centering
     \includegraphics[scale=0.15]{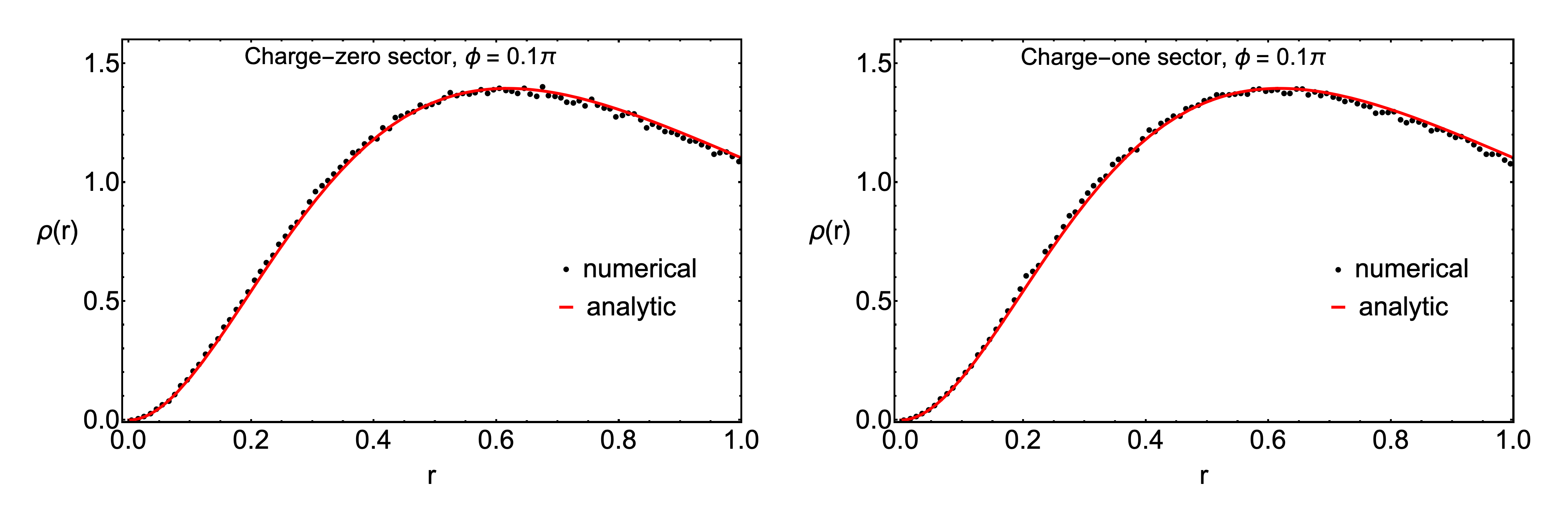}
     \includegraphics[scale=0.15]{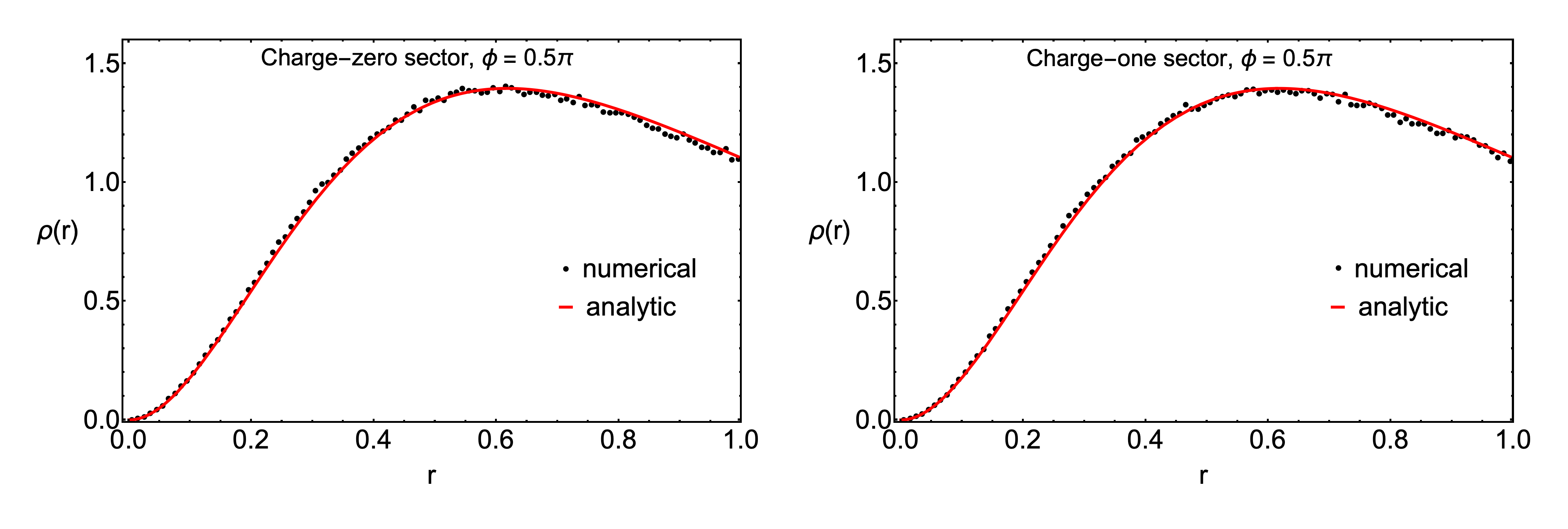}
     \caption{The distribution of spacing ratios in two charge sectors.  The black dots are obtained from numerically diagonalizing 100 realizations of $N=16$ Hamiltonians, and the red curves are the analytic formula \eqref{eqn:spacingRatioGUE}. }
     \label{fig:spacingRatioDistri}
 \end{figure}

We use exact diagonalization procedure to study the level statistics. 
For the sake of variety, we still study both the high and low energy regions of the spectrum, and we use the $p=1$ Hamiltonian 
\begin{equation}
H = \sum_{1\leq i_1<i_2 \leq N} J_{i_1 i_2}(T_{i_1}^+ T_{i_2}^- +T_{i_2}^+ T_{i_1}^- ),
\end{equation}
and the fluxes follow an i.i.d.  distribution  where $F_{ij} = \pm \phi$ with equal probabilities.  We also let $J_{i_1 i_2}$ be i.i.d. and binary-valued:
\begin{equation}
    J_{i_1 i_2} = \pm \sqrt{2\binom{N}{2}^{-1}}
\end{equation}
with equal probability.  We obtain the full spectrum of such Hamiltonians with $N=16$,  for two values of $\phi$ ($0.1\pi$ and $0.5\pi$). For each value of $\phi$ we compute the eigenvalues of 100 realizations of the Hamiltonians in given charge sectors.  In figure \ref{fig:spacingRatioDistri} we plot the results for charge-zero ($\binom{16}{8}=12870$  eigenvalues for each realization) and charge-one ($\binom{16}{7}=11440$  eigenvalues for each realization)  sectors.   We see that the results agree extremely well with the GUE result \eqref{eqn:spacingRatioGUE}, and hence the spectrum of our model exhibit RMT universality.
\begin{figure}[t!]
     \centering
     \includegraphics[scale=0.15]{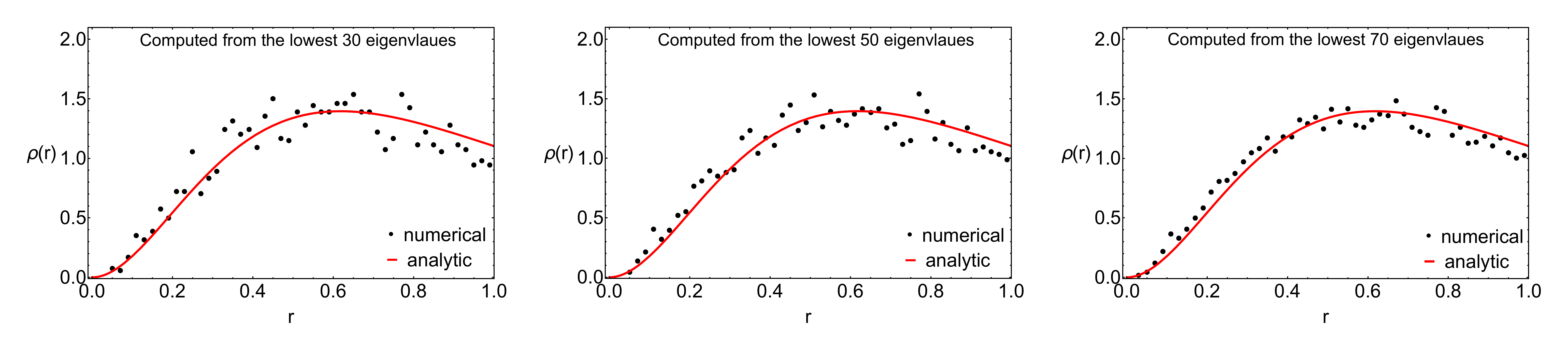}
     \caption{The distribution of spacing ratios for low-energy spectrum.  The numerical data involved are the same as those used in figure \ref{fig:spacingRatioDistri},  however here we only take the lowest eigenvalues of the charge-zero sector with $N=16, \phi =0.1\pi$.  The black dots are the numerical results and the red curves are the analytic formula \eqref{eqn:spacingRatioGUE}.}
     \label{fig:lowlyingRatios}
 \end{figure}
 
We can further ask if the low-energy part of spectrum exhibits RMT universality as well.  In figure \ref{fig:lowlyingRatios} we take the lowest eigenvalues of the charge-zero sector and plot the spacing ratio distributions.   Even when we take as few as the lowest 30 eigenvalues for each realization (out of the 12870 values),  $\rho(r)$ shows a clear resemblance to the GUE result $\rho_\text{GUE}(r)$,  although there are large fluctuations since we have fewer samples.   Convergence quickly improves as we increase the number of eigenvalues:  with 70 eigenvalues for each realization,  the convergence to $\rho_\text{GUE}(r)$ is already unmistakable.  We also expect the convergence to improve as we increase $N$ or the number of realizations.
The spacing ratio only captures the short-range correlations which are the ones relevant to RMT universality.  We expect the long-range correlations to be also present and deviate from RMT predictions.  Such long-range correlations are more visible in measures such as the connected part of spectral form  factor (as an early-time bump)  or the number variance (as a large tail) \cite{Jia_2020,Jia:2019orl}.

Since the spectrum exhibits RMT universality,   we expect our model to eigenstate-thermalize,  all the way down to  very low energies.
\section{Conclusion}
We have constructed a class of NCFT$_1$ models based on canonical bosons that are free of low-temperature orderings. The construction is based on the fairly general idea that a large amount of  uniform but random fluxes in the microcscopic Fock spaces achieve the following simultaneously:
\begin{enumerate}
    \item At early time, the correlation functions are described by chord diagram solutions, and therefore give rise to Schwarzian physics.
    \item At late time, the suppression of return amplitudes make wavefunctions very delocalized. This implies eigenstate thermalization and the lack of low-temperature orderings.
\end{enumerate}
In other words two universalities (Schwarzian and RMT) are achieved  simultaneously for such models, largely independent of the nature of the underlying Fock space. This explains why purely bosonic NCFT$_1$ models are so hard to build in the $p$-local approach (not enough random exchange phases), and why they do not present a challenge in our approach (random exchange phases are engineered in).  Moreover,  our approach allows us to use chord diagram combinatorics without taking a double scaling limit (though not forbidden either). This gives us some extra advantages compared to the previous double-scaled models that carry conserved charges, for example our construction has much better-behaved thermodynamic functions and subleading corrections to Lyapunov exponents. 

Clearly this picture allows for many more constructions of  NCFT$_1$ models beyond the ones presented in this paper.  For example it would be interesting to construct models with supersymmetries which have both elementary bosons and fermions. It would also be interesting if one can construct models whose charge scaling behaviors actually follow those of black holes.

\acknowledgments{
We have benefited from illuminating discussions with Micha Berkooz, Yingfei Gu, Antonio Garic\'{i}a-Garc\'{i}a, Matthew Heydeman,  Ziming Ji, Vladi Narovlansky,  Cheng Peng and Zhenbin Yang.  Part of the work was done when the YJ was visiting Shanghai Jiao Tong University,  Yau Mathematical Sciences Center (YMSC) at Tsinghua University and Institute for Advance Studies at Tsinghua University (IASTU), particularly thanks to the hospitality of Antonio Garic\'{i}a-Garc\'{i}a, Chi-Ming Chang, Zhenbin Yang, Wei Song,  Wen-Xin Lai, Xuyao Hu and Yixu Wang.}
\appendix
\section{Interpreting the formally  divergent integral \eqref{eqn:fixedchargemomIntegral1}}\label{app:formalIntegral}
Equation \eqref{eqn:fixedchargemomIntegral1} deals with a formally divergent integral
\[ \Tr\left(\delta(R-Q) H^k \right) \propto \int_{-\pi}^\pi da\left[\frac{e^{ia}}{(1-e^{ia})^2}\right]^{kp/2}   \exp\left[-iN\mathcal{Q} a -N\log(1-e^{ia})\right].\]
The particle number $R$ is a sum over non-negative operators $b_i^\dagger b_i$,  so the fixed charge condition enforces that the subspace we are tracing over is finite-dimensional, and therefore the left-hand side must be finite.   The divergence of the right-hand side comes from an inappropriate change of orders of taking limits.   For any give charge $Q$,  we only need to evaluate $\Tr (e^{i a R} H^k)$ up to a cut-off charge $R = \Lambda \geq Q$ and then Fourier-transform to get the exact answer.   In fact for simplicity we can do an overkill and demand that each individual $b^\dagger_i b_i$ is cut off at $\Lambda$. Then in place of the formally divergent (at $a=0$) factor $(1-e^{i a})^{-1}$ we should have got 
\begin{equation}
\sum_{n=0}^\Lambda e^{i n a} = \frac{1-e^{i a (\Lambda +1)}}{1-e^{ia}}
\end{equation}
which is regular everywhere.  If we use this expression to do the integral,  the result should be finite and independent of $\Lambda$ (as long as $\Lambda \geq Q$).    Since the charge scaling we are interested in is $Q = N \mathcal{Q}\sim N$,    $\Lambda$ must at least scale as $\sim N$.  We can now perform  the saddle-point evaluation of this completely well-defined integral.  The new  large $N$  exponent is 
\begin{equation}
-iN\mathcal{Q} a + N\log \frac{1-e^{ia (\Lambda+1)}}{1-e^{ia}}
\end{equation}
and the saddle point equation is 
\begin{equation}
\frac{1+(\Lambda e^{i a} - \Lambda-1)e^{i a \Lambda}}{(1-e^{-i a}) (e^{i a (\Lambda+1)}-1)} = \mathcal{Q}.
\end{equation}
Since the right-hand side is order-one in $N$ but $\Lambda$ grows faster than $N$,  we can divide the saddle points into two cases. The first case is when 
\begin{equation}
\Lambda e^{i a \Lambda} \to 0, \text{as } \Lambda \to \infty.
\end{equation}
This is the one that gives the original stable saddle $a = i \log (1+\mathcal{Q}^{-1})$ as $\Lambda$ goes to infinity.  One can check this saddle has a thimble (contour of constant phase that passes through the saddle point) that is homologous to a subinterval of the original integration domain $[-\pi, \pi]$.   The second type of solutions are highly dependent  on $\Lambda$ and produce saddle-point actions that are highly dependent on $\Lambda$.  However we have established that the exact value of the integral is independent of $\Lambda$, therefore it must be that they do not really contribute in the large $N$ limit.  It could be because these saddles are subleading,  or they give thimbles that are not homologous to any subinterval of $[-\pi, \pi]$.   One can numerically verify this is indeed the case.

To summarize,  the formally divergent integral \eqref{eqn:fixedchargemomIntegral1} really means a completely convergent one with a cutoff.   The formal divergence appears because we are not taking limits in the right order.  However, cutoff independence largely implies that the saddle that contributes must be stable as we vary the cutoff,  and this is why it is still correctly captured by the formally divergent version. So we would get the correct saddle-point value even if we did not notice the formal divergence in the first place.

\bibliographystyle{JHEP}
\bibliography{library}
\end{document}